\documentclass[12pt]{article}
\usepackage{amsmath}
\usepackage{graphicx,psfrag,epsf}
\usepackage{enumerate}
\usepackage{natbib}
\usepackage{hyperref}
\newcommand{\blind}{0}

\usepackage{graphics}

\usepackage{adjustbox}
\usepackage{amssymb}
\usepackage{bm}
\usepackage{graphicx}
\usepackage{natbib}
\usepackage{amsmath}
\makeatletter
\newcommand{\distas}[1]{\mathbin{\overset{#1}{\kern\z@\sim}}}%
\newsavebox{\mybox}\newsavebox{\mysim}
\newcommand{\distras}[1]{%
  \savebox{\mybox}{\hbox{\kern3pt$\scriptstyle#1$\kern3pt}}%
  \savebox{\mysim}{\hbox{$\sim$}}%
  \mathbin{\overset{#1}{\kern\z@\resizebox{\wd\mybox}{\ht\mysim}{$\sim$}}}%
}
\usepackage{breqn}
\usepackage{multirow}
\usepackage{bm}
\usepackage{multirow}
\usepackage[normalem]{ulem}
\useunder{\uline}{\ul}{}
\usepackage{graphicx}
\usepackage{wrapfig}
\usepackage{lscape}
\usepackage{rotating}
\usepackage{epstopdf}
\usepackage{comment}

\usepackage[utf8]{inputenc}
\usepackage[english]{babel}
 \usepackage{amsthm}

 \newtheorem{theorem}{Theorem}
\newtheorem{corollary}{Corollary}

\newtheorem{property}{Property}
\usepackage{graphicx}
\usepackage{subcaption}

\usepackage{booktabs,subcaption,amsfonts,dcolumn}

\usepackage{comment}

\begin{document}

\def\spacingset#1{\renewcommand{\baselinestretch}%
{#1}\small\normalsize} \spacingset{1}


\if0\blind
{
  \title{\bf Geostatistical Modeling of Positive Definite Matrices: An Application to Diffusion Tensor Imaging}

  \author{
  Zhou Lan\\
    Department of Statistics, North Carolina State University\\
    and \\
    Brian J Reich \\
    Department of Statistics, North Carolina State University\\
    and\\
      Joseph Guinness\\
      Department of Statistical Science, Cornell University\\
      and\\
       Dipankar Bandyopadhyay \\
      Department of Biostatistics, Virginia Commonwealth University\\
       and\\
    Liangsuo Ma \\
    Department of Radiology, Virginia Commonwealth University\\
    and \\
    F. Gerard Moeller \\
    Institute of Drug \& Alcohol Studies, Virginia Commonwealth University\\
    }

  \maketitle
} \fi

\if1\blind
{
  \bigskip
  \bigskip
  \bigskip
  \begin{center}
    {\LARGE\bf Title}
\end{center}
  \medskip
} \fi

\bigskip
\begin{abstract}
Geostatistical modeling for continuous point-referenced data has been extensively applied to neuroimaging because it produces efficient and valid statistical inference. However, diffusion tensor imaging (DTI), a neuroimaging characterizing the brain's anatomical structure, produces a positive definite matrix for each voxel. Currently, only a few geostatistical models for positive definite matrices have been proposed because introducing spatial dependence among positive definite matrices properly is challenging. In this paper, we use the spatial Wishart process, a spatial stochastic process (random field) where each positive definite matrix-variate marginally follows a Wishart distribution, and spatial dependence between random matrices is induced by latent Gaussian processes. This process is valid on an uncountable collection of spatial locations and is almost-surely continuous, leading to a reasonable means of modeling spatial dependence. Motivated by a DTI dataset of cocaine users, we propose a spatial matrix-variate regression model based on the spatial Wishart process. A problematic issue is that the spatial Wishart process has no closed-form density function. Hence, we propose an approximation method to obtain a feasible Cholesky decomposition model and show that the Cholesky decomposition model is asymptotically equivalent to the spatial Wishart process model. A local likelihood approximation method is also applied to achieve fast computation. The simulation studies and real data analysis demonstrate that the Cholesky decomposition process model produces reliable inference and improved performance compared to other methods.
\end{abstract}

\noindent%
{\it Keywords:}  Diffusion tensor imaging; Geostatistical modeling; Random matrix; Positive definite matrix; Spatial random fields; Spatial Wishart process; Cholesky decomposition process

\spacingset{1.5}

\section{Introduction}
\label{s:int}
Diffusion tensor imaging (DTI), a magnetic resonance imaging (MRI) technique, is used to measure the diffusion process of water molecules in the brain \citep{soares2013hitchhiker}. Estimated $3\times 3$ positive definite matrices summarize the water diffusion at each location in the brain. The positive definite matrix is also called a diffusion tensor (DT), representing the covariance of the local 3D Brownian motion \citep{schwartzman2006random,dryden2009non}. Since DTI has been used extensively to map white matter tractography in the brain, it has an advantage over other MRI-based techniques in revealing abnormal topological organization in the brain \citep{lo2010diffusion}. A primary clinical objective is to understand how covariates (e.g., age, gender, drug use) affect DTs, reflecting its effects on brain structure.

Incorporating spatial dependence is important for achieving efficient and valid inference in imaging data analysis \citep{spence2007accounting,wu2013mapping,xue2018bayesian}. Recently, \citet{lan2019spatial} also reveal that incorporating spatial dependence leads to improved performance in DTI region of difference selection, validated by an application to a cocaine user data-set \citep{ma2017preliminary}. In point-referred data, geostatistical modeling is a useful approach. It is a class of models based on continuous spatial variation, providing a smooth surface over locations (e.g., spatial Gaussian process model). Current geostatistical modeling only focuses on random variables following univariate or multivariate distributions (e.g., univariate/multivariate Gaussian, Poisson). However, the voxel-level variable in DTI is a positive definite matrix and only a few relevant works have been proposed for spatially-varying positive definite matrices \citep{gelfand2004nonstationary}. This triggers our study of geostatistical modeling of positive definite matrices.

Previous attempts to analyze DTI data can be broadly classified into univariate modeling and matrix-variate modeling. To avoid the complexity caused by matrix-variate data, univariate modeling projects a DT onto a descriptive scalar quantity such as the magnitude of isotropy, magnitude/fractional of anisotropy, or mode of anisotropy \citep{ennis2006orthogonal}. Among these scalar quantities, magnitude/fractional of anisotropy is the most popular \citep[see][]{lane2010diffusion,ma2017preliminary}. However, since these projections are surjective (e.g., different DTs may project onto the same scalar quantity), the loss of information caused by univariate modeling is irreversible. To this end, matrix-variate modeling has been proposed via parameterizing the DTs as matrix-variate random distributions such as Wishart distribution \citep{dryden2009non} and Gaussian random ellipsoid distribution \citep{schwartzman2008inference}. However, these matrix-variate models have not yet been extended to spatial modeling because incorporating spatial dependence for positive definite matrices is non-trivial.

To mitigate these issues, we propose a spatial matrix-variate regression model. The covariates are incorporated through the Cholesky decomposition \citep{zhu2009intrinsic}, and the coefficients are spatially-varying to capture local covariate effects. The spatial dependence among positive definite matrices is achieved by the spatial Wishart process, a spatial random field (stochastic process) supporting spatially dependent Wishart matrices. The first use of the spatial Wishart process was a prior for a spatially-varying covariance matrix \citep{gelfand2004nonstationary}. In this paper, the spatial Wishart process is used as a model for positive definite matrix observations. Considering that the literature comprehensively describing the statistical properties of the spatial Wishart process is sparse, we further prove that the spatial Wishart process as a random field on uncountable locations is valid and almost-surely continuous. Although the model based on the spatial Wishart process is elegant with several nice properties, a bottleneck of the spatial Wishart process is that its probability density function is intractable \citep{viraswami1991multivariate}. Therefore, instead of directly modeling the DTs, we propose a new Cholesky decomposition process model that approximates the original model by taking the Cholesky decomposition of positive definite matrices as the responses. The Cholesky decomposition process model is composed of six univariate spatial Gaussian processes, where the parameters retain the interpretations of the original model. Via theoretical results and simulation studies, we show that the Cholesky decomposition process model is an asymptotic approximation and useful working model. The theoretical results make an important contribution of both direct and potential value in applications of DTI and other fields, simplifying matrix-variate models relying on dependent Wishart matrices \citep[e.g.,][]{karagiannidis2003efficient,smith2007distribution,kuo2007joint} to multivariate models relying on Gaussian distributions. We also deal with massive spatial data by Vecchia's method \citep{vecchia1988estimation,datta2016hierarchical}, a local likelihood approximation that approximates the joint density of spatial variables as a product of conditional densities. To demonstrate our proposal, we further investigate its performance using simulation studies, and provide data analysis on cocaine user data \citep{ma2017preliminary}, in comparison to the univariate spatially-varying coefficient process model \citep{gelfand2003spatial}. To the best of our knowledge, this is the first work on exploring spatial associations in modeling positive definite matrix-variate data under the framework of geostatistical modeling, with some key theoretical contributions of multivariate analysis and applications to DTI. 

\section{Spatial Wishart Process Model}
\label{s:SWP}
A typical DTI data set \citep[e.g.,][]{ma2017preliminary} usually includes DTs from each subject $i\in \{1,2,...,N\}$ at each voxel $\bm{s}\in\{\bm{s}_1,...,\bm{s}_n\}$, and subject-level covariates (e.g., age, education level, medical treatments). The primary clinical objective is to detect local covariate effects on DTs. Let $\bm{A}_i(\bm{s})$ be the $3\times 3$ positive definite DT matrix of subject $i$ measured at voxel $\bm{s}$ and $\bm{X}_i$ be a design matrix containing an intercept and $d$ covariates. The positive definite DT matrices are modeled as parameterized Wishart matrices \citep{dryden2009non} to have mean matrix $\bm{\Sigma}_i(\bm{s})$ and degrees of freedom $m$, denoted as $\bm{A}_i(\bm{s})\sim \mathcal{W}(\bm{\Sigma}_i(\bm{s}), m)$. To model spatial dependence and ensure that $\bm{A}_i(\bm{s})$ is a positive definite matrix, we decompose $\bm{A}_i(\bm{s})$ as
\begin{equation}
\label{eq:decomp}
\bm{A}_i(\bm{s})=\bm{L}_i(\bm{s})\bm{U}_i(\bm{s})\bm{L}_i(\bm{s})^T.    
\end{equation}
In this decomposition, $\bm{U}_i(\bm{s})$ has mean $\bm{I}_3$ and is the spatially dependent \textit{residual term} modeling variation which cannot be explained by the covariates and the \textit{regression term} $\bm{L}_i(\bm{s})$ is the lower-triangle Cholesky matrix of the mean matrix $\bm{\Sigma}_i(\bm{s})$, i.e., $\mathbb{E}\bm{A}_i(\bm{s})=\bm{\Sigma}_i(\bm{s})=\bm{L}_i(\bm{s})\bm{L}_i(\bm{s})^T$. The spatial Wishart process for $\bm{U}_i(\bm{s})$ is described in Section \ref{s:res} and the regression construction for $\bm{L}_i(\bm{s})$ as a function of $\bm{X}_i$ is described in Section \ref{s:pop}. We refer to this model as spatial Wishart process model in the rest of the paper.

\subsection{\textit{Residual Term}: Spatial Wishart Process}
\label{s:res}
In this subsection, we introduce the spatial Wishart process as a means of modeling spatial dependence. \citet{gelfand2004nonstationary} provide the construction of the spatial Wishart process, which is stated as follows. For $j\in\{1,2,...,m\}$, let $\{\bm{Z}_j(\bm{s}):\bm{s}\in\mathcal{D}\}$ be a mean-zero $p$-dimensional multivariate Gaussian process with $p\times p$ cross-covariance matrix $\bm{\Sigma}$ and spatial dependence function $\mathcal{K}(\bm{s},\bm{s}'\mid \bm{\Phi})$ determined by parameters $\bm{\Phi}$, i.e., $cov(\bm{Z}_j(\bm{s}),\bm{Z}_j(\bm{s}'))=\mathcal{K}(\bm{s},\bm{s}'\mid \bm{\Phi})\times\bm{\Sigma}$, denoted as $\bm{Z}_{j}\sim \mathcal{GP}(\bm{0},\mathcal{K}(\bm{s},\bm{s}'\mid \bm{\Phi}), \bm{\Sigma})$. If for each $\bm{s}\in\mathcal{D}$, we collect $\bm{U}(\bm{s})=\sum_{j=1}^m\bm{Z}_{j}(\bm{s})\bm{Z}_{j}^T(\bm{s})/m\sim\mathcal{W}(\bm{\Sigma},m)$, then the collection $\{\bm{U}(\bm{s}):\bm{s}\in\mathcal{D}\}$ is a spatial Wishart process, a random field supporting spatially dependent Wishart matrices. The spatial Wishart process can be understood as a two-level hierarchical model where the spatial dependence of Wishart matrix $\bm{U}(\bm{s})$ is induced by the latent spatial Gaussian processes $\{\bm{Z}_{j}\}$. Also, in light of the application to DTI, we assume $p=3$ in default.

In applications, the number of locations in $\mathcal{D}$ is usually finite. However, \citet{gelfand2010continuous} emphasize the importance to ensure a valid mathematical specification of a spatial stochastic process. Thus, in Theorem \ref{thm:swp}, we use the Kolmogorov extension theorem \citep{oksendal2003stochastic} to prove that it is a valid stochastic process if $\mathcal{D}$ is an uncountable collection of spatial locations (Appendix I). 
\begin{theorem}[Spatial Wishart Process]
\label{thm:swp} The spatial Wishart process $\{\bm{U}(\bm{s}):\bm{s}\in\mathcal{D}\}$ is a valid stochastic process (random field).
\end{theorem}

Based on the fact that the random field is valid, we can also show that this field is almost-surely continuous (Property \ref{prop:conti}). The proof of almost-sure continuity is based on \citet{kent1989continuity} (Appendix I).
\begin{property}[Almost-Sure Continuity]
\label{prop:conti}
Let $\{\bm{U}(\bm{s}):\bm{s}\in\mathcal{D}\}$ be a spatial Wishart process. If the correlation function $\mathcal{K}(\bm{s},\bm{s}'\mid \bm{\Phi})$ has a second-order Taylor series expansion with remainder that goes to $0$ at a rate of $2+\delta$ for some $\delta>0$, $\bm{U}(\bm{s})$ converges weakly to $\bm{U}(\bm{s}_0)$ with
probability one as $\mid \mid \bm{s}-\bm{s}_0\mid \mid \rightarrow 0$.
\end{property}

Considering that neuroimaging data is usually is collected at a high resolution and the disease status at proximally-located/neighboring voxels can be similar \citep[see][]{wu2013mapping,xue2018bayesian}, the residuals $\bm{U}_i(\bm{s})$ should be smooth and spatially dependent. Therefore, we model the residuals $\{\bm{U}_i(\bm{s}):\bm{s}\in \mathcal{D} \}$ for $i\in\{1,2,...,N\}$ as realizations of a spatial Wishart process with degrees of freedom $m$, cross-covariance matrix $\bm{I}$, and correlation function $\mathcal{K}(\bm{s},\bm{s}'\mid \bm{\Phi})$, denoted as \begin{equation}
    \bm{U}_i\sim \mathcal{SWP}(m,\mathcal{K}(\bm{s},\bm{s}'\mid \bm{\Phi}), \bm{I}).
\end{equation}
Setting the cross-covariance matrix to $\bm{I}$ preserves the designed marginal distribution $\bm{A}_i(\bm{s})\sim \mathcal{W}(\bm{\Sigma}_i(\bm{s}),m)$.

In spatial statistics and neuroimaging, understanding the spatial dependence is essential. We quantify spatial dependence of the spatial Wishart process using the expected squared Frobenius norm $\mathcal{V}(\bm{s},\bm{s}')=\mathbb{E}\mid \mid \bm{U}(\bm{s})-\bm{U}(\bm{s}')\mid \mid _F^2$, where $\mid \mid .\mid \mid _F$ is the Frobenius norm. The expected squared Frobenius norm can also be understood as a generalized variogram \citep{cressie1992statistics} for matrix-variate data \citep{lan2019spatial}, where an increasing spatial dependence of $\bm{U}(\bm{s})$ and $\bm{U}(\bm{s}')$ leads to a smaller $\mathcal{V}(\bm{s},\bm{s}')$. Through the variogram, we find that the spatial Wishart process $\mathcal{SWP}(m,\mathcal{K}(\bm{s},\bm{s}'\mid \bm{\Phi}),\bm{\Sigma})$ is separable \citep{cressie1992statistics} since
\begin{equation}
    \mathcal{V}(\bm{s},\bm{s}')=\gamma(m,\bm{\Sigma}) [1-\mathcal{K}(\bm{s},\bm{s}'\mid \bm{\Phi})^2],
\end{equation}
where the term $1-\mathcal{K}(\bm{s},\bm{s}'\mid \bm{\Phi})^2$ is the spatial term and $\gamma(m,\bm{\Sigma})=\frac{2}{m}Tr(\bm{\Sigma\Sigma})+\frac{2}{m} Tr(\bm{\Sigma})Tr(\bm{\Sigma})$ is the non-spatial term. The property of spatial separability  makes the residual variation more transparent: The cross-dependence of positive definite matrices depends on $\gamma(m,\bm{\Sigma})$ such that an increasing $m$, primarily controlling the variance of a Wishart matrix, leads to smaller cross-dependence; A larger spatial correlation of the underlying Gaussian processes leads to larger spatial dependence.

\begin{figure}[t!]
    \centering
    \begin{subfigure}[t!]{1\textwidth}
       \includegraphics[width=\textwidth]{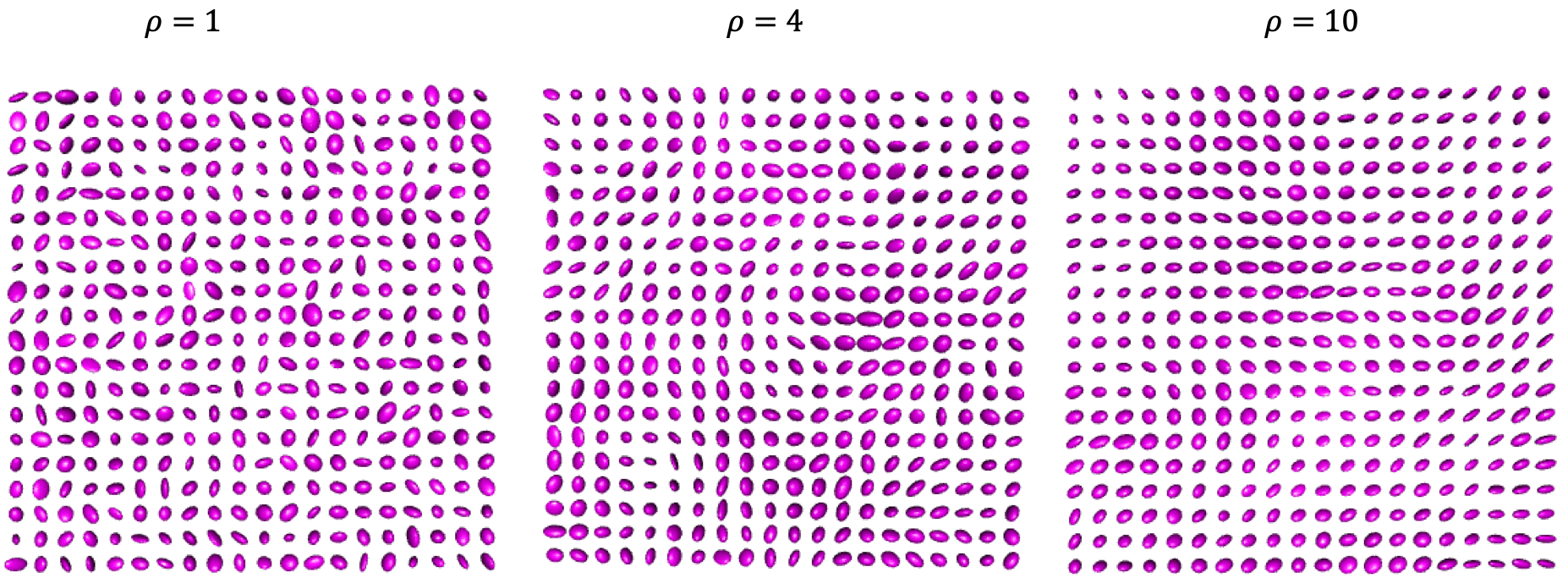}
          \caption{Simulated positive definite matrices with $m=10$.}\label{fig:sim1}
    \end{subfigure}
     \begin{subfigure}[t!]{1\textwidth}
    \includegraphics[width=\textwidth]{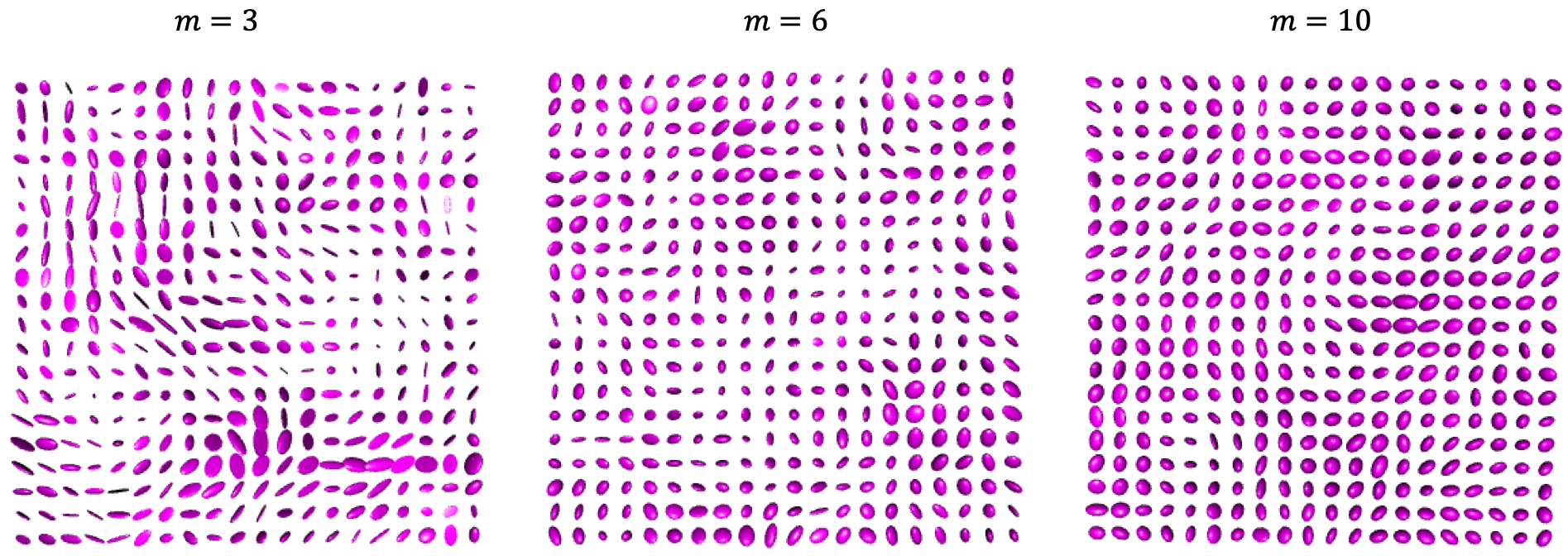}
    \caption{Simulated positive definite matrices with $\rho=4$.}\label{fig:sim2}
    \end{subfigure}
    \caption{Simulated positive definite matrices from standard spatial Wishart processes, i.e., with mean equal to the identity matrix. The spatial dependence of positive definite matrices depends on the range parameter $\rho$. The cross-dependence of positive definite matrices depends on the degrees of freedom $m$.}\label{fig:swp}
\end{figure}

The spatial dependence can be visualized via realizations of the standard spatial Wishart processes ($\bm{\Sigma}=\bm{I}$) on a $20\times 20$ grid with spacing of 1 between adjacent grid points. Given the spatial correlation function is exponential  $\mathcal{K}(\bm{s},\bm{s}'\mid \rho)=\exp\left[-\frac{\mid \mid \bm{s}-\bm{s}'\mid \mid }{\rho}\right]$, we visualize the positive definite matrices in two dimensions as ellipsoids in Figure \ref{fig:swp}. In Figure \ref{fig:sim1}, the positive definite matrices are simulated with $m=3$ and $\rho=1,4,10$, where a larger $\rho$ leads to stronger spatial dependence; In Figure \ref{fig:sim2}, positive definite matrices are simulated with $\rho=4$ and $m=3,6,10$, where a larger $m$ leads smaller cross-dependence. Since the three cases in Figure \ref{fig:sim2} maintain the same level of spatial dependence, we may also identify that the spatial and non-spatial variations are separable.

\subsection{\textit{Regression Term}: Cholesky Decomposition}
\label{s:pop}
Expressing the mean matrix $\bm{\Sigma}_i(\bm{s})$ in terms of $\bm{X}_i$ is not straightforward \citep{zhu2009intrinsic,yuan2012local} because the responses $\bm{A}_i(\bm{s})$ are in a Remannian manifold but the covariates $\bm{X}_i$ are in Euclidean space. Following \citet{zhu2009intrinsic}, we regress the $(k,l)$-th element of $\bm{L}_i(\bm{s})$, denoted as $l_{ikl}(\bm{s})$ on $\bm{X}_i$ as
\begin{equation}
\label{eq:regss}
\begin{aligned}
\log l_{ikk}(\bm{s})=\bm{X}_i\bm{\beta}_{kk}(\bm{s}), \quad l_{ikl}(\bm{s})=\bm{X}_i\bm{\beta}_{kl}(\bm{s})\quad \text{for $k>l$},
\end{aligned}
\end{equation}
where $\bm{\beta}_{kl}(\bm{s})=[\beta_{0kl}(\bm{s}), \beta_{1kl}(\bm{s}), ..., \beta_{dkl}(\bm{s})]^T$ is the spatially-varying coefficient vector and $ \beta_{jkl}(\bm{s})$ is the coefficient associated with the $j$-th covariate. The roles of the coefficients $\bm{\beta}_{kl}(\bm{s})$ can be explained as linear effect on $\log l_{ikk}(\bm{s})$ or $l_{ikl}(\bm{s})$. To model the spatial dependence of the mean effect, we assign a mean-zero spatial Gaussian process prior on $\bm{\beta}(\bm{s})=[\bm{\beta}_{11}(\bm{s})^T, \bm{\beta}_{22}(\bm{s})^T, \bm{\beta}_{33}(\bm{s})^T, \bm{\beta}_{21}(\bm{s})^T, \bm{\beta}_{31}(\bm{s})^T, \bm{\beta}_{32}(\bm{s})^T]^T$, denoted as $\bm{\beta}\sim\mathcal{GP}(\bm{0},\mathcal{K}(\bm{s},\bm{s}'\mid \bm{\Phi}_{\beta}),\sigma_{\beta}^2\bm{I})$, where $\bm{\Phi}_{\beta}$ is a set of spatial parameters controlling the spatial dependence of mean process, and $\sigma_{\beta}^2$ is the variance of the Gaussian process.
 
  \section{Cholesky Decomposition Process Model}
 \label{s:working}
 \citet{viraswami1991multivariate} and others \citep[e.g.,][]{blumenson1963properties,smith2007distribution} show that a closed-form probability density function of the spatial Wishart process model is available if the latent spatial precision matrix is tri-diagonal. However, the joint probability density function of multiple locations is not appropriate to be used for large-scale spatial modeling because this assumption of tri-diagonal precision matrix is unrealistic in spatial modeling. Therefore, to approximate the spatial Wishart process model, we further propose the Cholesky decomposition process model. The Cholesky decomposition process model is specified on the Cholesky decomposition elements of $\bm{A}_i(\bm{s})$, denoted as $\{t_{ikl}(\bm{s}):k\geq l,\bm{s}\in \mathcal{D}\}$. The model is
 
 \begin{equation}
 \label{eq:CDP}
     \begin{aligned}
     \textbf{Diagonal:}\ & \sqrt{2}\log t_{ikk}\sim \mathcal{GP}\left(\sqrt{2}\bm{X}_i\bm{\beta}_{kk}(\bm{s}),\mathcal{C}(\bm{s},\bm{s}'\mid \bm{\Phi}_u),\sigma_m^2\right)&\quad &\text{for $k=1,2,3$}, \\
     \textbf{Off-Diagonal:}\ & t_{{ikl}}\mid \bar{t}_{ikk}\sim \mathcal{GP}(\bm{X}_i\bm{\beta}_{kl}(\bm{s}),\mathcal{C}(\bm{s},\bm{s}'\mid \bm{\Phi}_u)\bar{t}_{ikk}(\bm{s})\bar{t}_{ikk}(\bm{s}'),\sigma_m^2)&\quad &\text{for $k>l$},\\
     \end{aligned}
 \end{equation}
In this  expression, $\sqrt{2}\bm{X}_i\bm{\beta}_{kk}(\bm{s})$ and $\bm{X}_i\bm{\beta}_{kl}(\bm{s})$ are marginal means of the diagonal and off-diagonal Gaussian processes at location $\bm{s}$, respectively. Also, $\bar{t}_{ikk}(\bm{s})=\exp(\bm{X}_i\hat{\bm{\beta}}_{kk}(\bm{s}))$ and $\hat{\bm{\beta}}_{kk}(\bm{s})$ is the ordinary least squares estimates computed using only data at voxel $\bm{s}$ from regressing $\log t_{ikk}(\bm{s})$ on $\bm{X}_i$.

 To provide a rigorous mathematical validation, we also prove that $\{\bm{A}_i(\bm{s})=\bm{T}_i(\bm{s})\bm{T}_i(\bm{s})^T:\bm{s}\in\mathcal{D}\}$ is a valid stochastic process and almost-surely continuous (Theorem \ref{thm:valid}), where $\bm{T}_i(\bm{s})$ is the lower-triangle Cholesky matrix.
  \begin{theorem}[Cholesky Decomposition Process]
 \label{thm:valid}
$\{\bm{A}_i(\bm{s})=\bm{T}_i(\bm{s})\bm{T}_i(\bm{s})^T:\bm{s}\in\mathcal{D}\}$ is a valid stochastic process. Also, if the correlation function $\mathcal{C}(\bm{s},\bm{s}'\mid \bm{\Phi})$ has a second-order Taylor series expansion with remainder that goes to $0$ at a rate of $2+\delta$ for some $\delta>0$, $\bm{A}(\bm{s})$ converges weakly to $\bm{A}(\bm{s}_0)$ with
probability one as $\mid \mid \bm{s}-\bm{s}_0\mid \mid \rightarrow 0$.
\end{theorem}

 To link to the spatial Wishart process model, we assume that $\mathcal{C}(\bm{s},\bm{s}'\mid \bm{\Phi}_u)=\mathcal{K}(\bm{s},\bm{s}'\mid \bm{\Phi}_u)^2$ and $\sigma_m^2=\frac{1}{m}$. Given the asymptotic properties in Theorem \ref{thm:eq}, we may conclude that asymptotically the two models are equivalent and the parameters in the two models have the same interpreations: $\bm{\beta}_{kk}(\bm{s})$ controls the mean of $\log t_{ikk}(\bm{s})$ and partially describes local variation of $t_{ikl}(\bm{s})$; $\bm{\beta}_{kl}(\bm{s})$ controls the mean of $t_{ikl}(\bm{s})$; $\bm{\Phi}_u$ controls the spatial residual dependence. Furthermore, if we modify that $\bar{t}_{ikk}(\bm{s})=\exp(\bm{X}_i{\bm{\beta}}_{kk}(\bm{s}))$ in (\ref{eq:CDP}), the condition that $N\rightarrow\infty$ can be omitted to show that $\{\sqrt{m}[e_{ikl}(\bm{s}_1), ...,  e_{ikl}(\bm{s}_n)]^T\mid \bm{S}_{e_{ikl}} \}$ converges in distribution to $\{\sqrt{m}[t_{ikl}(\bm{s}_1), ..., t_{ikl}(\bm{s}_n)]^T\mid \bar{t}_{ikk}\}$. However, we show that the specification in (\ref{eq:CDP}) leads to computationally efficient Gibbs sampling for coefficients (Section \ref{sec:computing}) and a reasonable trade-off according to the simulation results showing the closeness of parameter estimation (Section \ref{s:sim}). 
 
 \begin{theorem}[Asymptotic Properties]
 \label{thm:eq}
 For $i\in\{1,2,...,N\}$, let $\{t_{ikl}(\bm{s}):k\geq l,\bm{s}\in \mathcal{D}\}$ and $\{e_{ikl}(\bm{s}):k \geq l,\bm{s}\in \mathcal{D}\}$ be Cholesky decomposition elements of $\{\bm{A}_i(\bm{s}):\bm{s}\in \mathcal{D}\}$ following the Cholesky decomposition process model and the spatial Wishart process model, respectively. For $\bm{s}_1,...,\bm{s}_n\in\mathcal{D}$, we have the following asymptotic results:
 \begin{itemize}
     \item \textbf{Diagonal:}  As $m\rightarrow\infty$, $\sqrt{m}[\log e_{ikk}(\bm{s}_1)-\bm{X}_i\bm{\beta}_{kk}(\bm{s}_1), ..., \log e_{ikk}(\bm{s}_n)-\bm{X}_i\bm{\beta}_{kk}(\bm{s}_n)]^T$ converges in distribution to  $\sqrt{m}[\log t_{ikk}(\bm{s}_1)-\bm{X}_i\bm{\beta}_{kk}(\bm{s}_1), ..., \log t_{ikk}(\bm{s}_n)-\bm{X}_i\bm{\beta}_{kk}(\bm{s}_n)]^T$
      for $k=1,2,3$;\\
     \item \textbf{Off-Diagonal:} 
      \begin{itemize}
     \item Let $\bm{\mu}_{m,kl}$ (dependent of $m$) and $\bm{\mu}_{kl}$ (independent of $m$) be the means of $\{[e_{ikl}(\bm{s}_1), ...,  e_{ikl}(\bm{s}_n)]^T|\bm{S}_{e_{ikl}} \}$ and $\{[t_{ikl}(\bm{s}_1), ...,  t_{ikl}(\bm{s}_n)]^T|\bar{t}_{ikk} \}$, respectively. As $m\rightarrow\infty$, $\bm{\mu}_{m,kl}$ converges in probability to $\bm{\mu}_{kl}$, and $cor(e_{{ikl}}(\bm{s}),e_{{ikl}}(\bm{s}')\mid \bm{S}_{e_{ikl}})$ converges in probability to $cor(t_{{ikl}}(\bm{s}),t_{{ikl}}(\bm{s}')\mid \bar{t}_{ikk})=\mathcal{K}(\bm{s},\bm{s}'\mid \bm{\Phi})^2=\mathcal{C}(\bm{s},\bm{s}'\mid \bm{\Phi})$;
      \item If $\bm{\beta}_{kl}(\bm{s})=\bm{0}$ for all $\bm{s}\in\mathcal{D}$ and $k>l$, then as $m\rightarrow\infty$ and $N\rightarrow\infty$, $\{\sqrt{m}[e_{ikl}(\bm{s}_1), ...,  e_{ikl}(\bm{s}_n)]^T\mid \bm{S}_{e_{ikl}} \}$ converges in distribution to $\{\sqrt{m}[t_{ikl}(\bm{s}_1), ..., t_{ikl}(\bm{s}_n)]^T\mid \bar{t}_{ikk}\}$,
     for $k>l$;
     \item If $\bm{\beta}_{kl}(\bm{s})=\bm{0}$ for all $\bm{s}\in\mathcal{D}$ and $k>l$, and that $\bar{t}_{ikk}(\bm{s})=\exp(\bm{X}_i{\bm{\beta}}_{kk}(\bm{s}))$, then as $m\rightarrow\infty$, $\{\sqrt{m}[e_{ikl}(\bm{s}_1), ...,  e_{ikl}(\bm{s}_n)]^T\mid \bm{S}_{e_{ikl}} \}$ converges in distribution to $\{\sqrt{m}[t_{ikl}(\bm{s}_1), ..., t_{ikl}(\bm{s}_n)]^T\mid \bar{t}_{ikk}\}$,
     for $k>l$;
       \end{itemize}
 \end{itemize}
 $\bm{S}_{e_{ikl}}(\bm{s})$ is a partition of the latent Gaussian processes defined in spatial Wishart process i.e., $\{\bm{Z}_{ij}(\bm{s})=[{Z}_{ij1}(\bm{s}),{Z}_{ij2}(\bm{s}),{Z}_{ij3}(\bm{s})]^T:\bm{s}\in\mathcal{D}\}$ for $j\in\{1,2,...,m\}$ ($i$ is a fixed and given index here. The partition is to take the dimensions ``above" of $e_{ikl}$ such as $\{[{Z}_{ij1}(\bm{s}),...,{Z}_{ij(k-1)}(\bm{s})]^T:\bm{s}\in\mathcal{D}\}$ for all $j$. For example, $\bm{S}_{e_{i31}}$ or $\bm{S}_{e_{i32}}$ is $\{[{Z}_{ij1}(\bm{s}),{Z}_{ij2}(\bm{s})]^T:\bm{s}\in\mathcal{D}\}$ for all $j$, and $\bm{S}_{e_{i21}}(\bm{s})$ is $\{{Z}_{ij1}(\bm{s}):\bm{s}\in\mathcal{D}\}$ for all $j$.
 \end{theorem}

The asymptotic results apply for large degrees of freedom, $m$. In both models, large m corresponds to small residual variability, i.e., images with small noise. This is a reasonable condition in our motivating data (see Section \ref{s:app}) where the estimated residual variance is small. In comparison to the spatial Wishart process model, the Cholesky decomposition process model is a more computational convenient Gaussian processes. Moreover, since the underlying mechanism of the DT's spatial dependence is unknown, both models can be treated as proposed geostatistical models for DTI. All the proofs for the results in this section are summarized in Appendix II.

  \subsection{Computational Details}
 \label{sec:computing}
In this subsection, we give the computational details of this model. We fit the model using Markov chain Monte Carlo and assign weakly informative priors to parameters. Given that $\mathcal{K}$ is the Matern correlation function, we define $\rho_u,\nu_u\in\bm{\Phi}_u$ as the range and smoothness parameter of the residual dependence, and $\rho_{\beta},\nu_{\beta}\in\bm{\Phi}_{\beta}$ as the range and smoothness parameter of the mean dependence. We give priors to these parameters: $\log\rho_u$ and $\log\rho_{\beta}$ follow a normal distribution with mean $0$ and standard deviation $1$; $\log\nu_u$ and $\log\nu_{\beta}$ follow a normal distribution with mean $-1$ and standard deviation $1$; $\sigma_{\beta}^{-2}$ and $\sigma_m^{-2}$ follow a gamma distribution with shape $0.01$ and rate $0.01$, which are conjugate priors allowing Gibbs sampling. The coefficients $\bm{\beta}$ are also updated using Gibbs sampling because their full conditional distributions are Gaussian distributions.

The computational bottleneck of the Cholesky decomposition process model is factoring the large $n\times n$ covariance matrix of the residual dependence and mean dependence, known as the $\mathcal{O}(n^3)$ problem in spatial statistics \citep[e.g.,][]{heaton2018case}. We address this problem using Vecchia's method \citep{vecchia1988estimation}, a local likelihood approximation that approximates the joint density of spatial variables as a product of conditional densities. Let $\omega$ be an arbitrary Gaussian process. The approximate joint density is $p[w(\bm{s}_1), ..., w(\bm{s}_n)]=\prod_{i=1}^n p[w(\bm{s}_i)\mid w(\bm{s}_k),\bm{s}_k\in N(\bm{s}_i)]$, where $N(\bm{s}_i)$ is a set of neighboring locations of $\bm{s}_i$ \citep{datta2016hierarchical}. This reduces the computational complexity from $\mathcal{O}(n^3)$ to $\mathcal{O}(nq^3)$, where $q\ll n$ is the largest size of $N(\bm{s})$. This approximation is implemented for $t_{ikl}$, $\log t_{ikk}$, and $\bm{\beta}$, where lexicographical order of locations on the regular spatial grid is used and $N(\bm{s}_i)$ is the following $q$ locations with larger ranks. A sensitivity analysis is presented in Section \ref{s:sim} to investigate the impact of the tuning parameter $q$ on the Cholesky decomposition process model. 

\section{Simulation}
\label{s:sim}

In this section, we first investigate the performance of the Cholesky decomposition process model under data generated from either the spatial Wishart process or Cholesky decomposition process model, demonstrating that the Cholesky decomposition process model produces reliable results under different geostatistical settings. Also, since we apply Vecchia's approximation for fast computation, we conduct a sensitivity analysis to investigate the impact of $q$ on parameter estimation.

For both models, we generate the synthetic DTs on $20\times 20$ grids with spacing of 1 between adjacent grid points. To mimic a real DTI study, $N=10$ subjects are simulated with drug-use indicator $x_{i,drug}\in\{0,1\}$ and normalized age $x_{i,age}\in\mathbb{R}^+$. The simulation study involves 50 replications. For each replication, there are $5$ drug users ($x_{i,drug}=1$) and $5$ non-drug users ($x_{i,drug}=0$), and $x_{i,age}$ is generated by a positive half-normal distribution \citep{leone1961folded} with mean $0$ and variance $1$. For each replication, all the coefficients $\bm{\beta}$ are generated from a spatial Gaussian process with variance $\sigma_{\beta}=0.1$ and correlation function $\mathcal{K}(\bm{s},\bm{s}'\mid \bm{\Phi}_\beta)$. The Gaussian process mean for three covariates (Table \ref{tab:coeff}) simulates a scenario that drug has an effect on certain regions of the brain and increasing age may affect the whole brain. In all replications, we simulate the data with $\rho_u=\rho_{\beta}=2$, $\nu_{u}=\nu_{\beta}=0.5$, and $m=50$. To investigate if Vecchia's approximation with different $q$ affects the model performance, we set $q=10,50$ and compare it to the model without Vecchia's approximation. For each replication, we collect 5,000 Markov chain Monte Carlo samples after discarding 2,000 samples as burn-in.

\begin{table}[t!]
\centering
\caption{The spatial Gaussian process mean of the six coefficient vector for three covariates are summarized. $\mathcal{S}$ is a set of spatial locations inside a $4 \times 4$ region in the middle of the image.}\label{tab:coeff}
\begin{tabular}{ccc}
Covariate & Diagonal & Off-Diagonal  \\ 
Intercept & $\beta_{int,kk}=0,\ \forall \bm{s}$ & $\beta_{int,kk}=0,\ \forall \bm{s}$  \\
& & \\
 $x_{i,drug}$& \begin{tabular}[c]{@{}c@{}}$\beta_{drug,kk}=0.5,\ \text{for} \bm{s}\in \mathcal{S}$\\ $\beta_{drug,kk}=0,\ \text{for} \bm{s}\notin \mathcal{S}$\end{tabular} & $\beta_{drug,kl}=0,\ \forall \bm{s}$  \\
 & & \\
 $x_{i,age}$ & $\beta_{kk,age}=0.25,\ \forall \bm{s}$ & $\beta_{kl,age}=0.25,\ \forall \bm{s}$  \\ 
\end{tabular}
\end{table}

The simulation results in terms of mean absolute deviation of posterior mean estimates\footnote{$\mid \mathbb{E}[\theta\mid .]-\theta\mid $ where $\theta$ is the true value and $\mathbb{E}[\theta\mid .]$ is the posterior mean.}, $95\%$ posterior coverage\footnote{Empirical percentage that the true value is in the $95\%$ posterior}, and Monte Carlo standard deviation\footnote{$\sqrt{\frac{1}{T}\sum_{t=1}^T(\theta^{(t)}-\mathbb{E}[\theta\mid .])^2}$ where $\theta^{(t)}$ is the $t$-th Markov chain Monte Carlo sample and there are totally $T$ Markov chain Monte Carlo samples.} are summarized in Tables \ref{t:sim1} and \ref{t:sim2}. To have a concise presentation, the values about coefficient estimates (Table \ref{t:sim1}) are averaged over replications, voxels ($n$), and covariates ($d$). From the simulation result, we find that Vecchia's approximation is acceptable since the computational times are 6 hours, 11 hours, and 35 hours for models with 10 neighbors, 50 neighbors, and without Vecchia's approximation and the  mean absolute deviation is nearly identical for all the three methods. We further conduct simulations with $m=3$ (Table \ref{t:sim1_non} and \ref{t:sim2_non}). When the degrees of freedom $m$ are large, the theoretical asymptotic results (Theorem \ref{thm:eq}) apply because the parameter estimations are close. Otherwise, when the degrees of freedom $m$ are small, the theoretical asymptotic results (Theorem \ref{thm:eq}) does not apply and inflated coverage is found when data is generated from the SWP model.

\begin{table}[t!]
\caption{Asymptotic ($m=50$) simulation results for spatially-varying coefficients with the data generated from the Cholesky decomposition process model or the spatial Wishart process model. The results are summarized in terms of mean absolute deviation of posterior mean estimates, $95\%$ posterior coverage, and Monte Carlo standard deviation. The values are averaged over replications, voxels ($n$), and covariates ($d$).}\label{t:sim1}
\centering
\begin{tabular}{cccccccc}
\multirow{2}{*}{Parameter} & \multirow{2}{*}{$q$} & \multicolumn{2}{c}{MAD} & \multicolumn{2}{c}{Coverage $95\%$}& \multicolumn{2}{c}{MCSD}\\ 
 &  & CDP & SWP & CDP & SWP & CDP & SWP \\ 
\multirow{3}{*}{$\bm{\beta}_{11}$} & 10 & 0.104 & 0.104 & $95\%$ & $93\%$ &0.05&0.05 \\
 & 50 & 0.104 & 0.104 & $95\%$ & $95\%$ &0.05&0.05\\
 & Standard & 0.104 & 0.094 & $95\%$ & $96\%$&0.05&0.06 \\ 
\multirow{3}{*}{$\bm{\beta}_{22}$} & 10 & 0.103 & 0.103 & $94\%$ & $95\%$ &0.05&0.05\\
 & 50 & 0.103 & 0.103 & $94\%$ & $95\%$&0.05&0.06 \\
 & Standard & 0.104 & 0.092 & $91\%$ & $95\%$&0.05&0.05 \\ 
\multirow{3}{*}{$\bm{\beta}_{33}$} & 10 & 0.104 & 0.104 & $94\%$ & $94\%$ &0.05&0.05\\
 & 50 & 0.105 & 0.104 & $93\%$ & $96\%$ &0.05&0.05\\
 & Standard & 0.105 & 0.092 & $93\%$ & $97\%$ &0.05&0.06\\ 
\multirow{3}{*}{$\bm{\beta}_{21}$} & 10 & 0.107 & 0.097 & $97\%$ & $99\%$ &0.08&0.07\\
 & 50 & 0.110 & 0.097 & $97\%$ & $97\%$  &0.08&0.08\\
 & Standard & 0.107 & 0.096 & $93\%$ & $97\%$  &0.07&0.08\\ 
\multirow{3}{*}{$\bm{\beta}_{31}$} & 10 & 0.109 & 0.097 & $95\%$ & $99\%$ &0.07&0.07\\
 & 50 & 0.110 & 0.096 & $95\%$ & $95\%$ &0.07&0.08\\
 & Standard & 0.109 & 0.096 & $95\%$ & $99\%$ &0.07&0.08\\ 
\multirow{3}{*}{$\bm{\beta}_{32}$} & 10 & 0.109 & 0.097 & $95\%$ & $99\%$ &0.08&0.08\\
 & 50 & 0.111 & 0.097 & $95\%$ & $99\%$ &0.08&0.08\\
 & Standard & 0.109 & 0.097 & $95\%$ & $97\%$ &0.07&0.08\\ 
  \multicolumn{8}{l}{MAD: Mean Absolute Deviation;}\\
  \multicolumn{8}{l}{MCSD: Monte Carlo Standard Deviation;}\\
 \multicolumn{8}{l}{SWP: Spatial Wishart Process Model;}\\
 \multicolumn{8}{l}{CDP: Cholesky Decomposition Process Model.}
\end{tabular}
\end{table}

\begin{table}[t!]
\caption{Asymptotic ($m=50$) simulation results for spatial parameters with the data generated from the Cholesky decomposition process model or the spatial Wishart process model. The results are summarized in terms of mean absolute deviation of posterior mean estimates, $95\%$ posterior coverage, and Monte Carlo standard deviation. The values are averaged over replications.}\label{t:sim2}
\centering
\begin{tabular}{cccccccc}
\multirow{2}{*}{Parameter} & \multirow{2}{*}{$q$} & \multicolumn{2}{c}{MAD} & \multicolumn{2}{c}{Coverage $95\%$}& \multicolumn{2}{c}{MCSD}\\ 
 &  & CDP & SWP & CDP & SWP & CDP & SWP \\ 
\multirow{3}{*}{$\rho_u=2$} & 10 & 0.17 & 0.20 & $98\%$ & $98\%$ &0.15 & 0.19\\
 & 50 & 0.17 & 0.20 & $96\%$ & $96\%$ &0.16 & 0.19\\
 & Standard & 0.10 & 0.14 & $98\%$ & $96\%$ &0.16 & 0.20\\ 
\multirow{3}{*}{$\nu_u=0.5$} & 10 & 0.033 & 0.033 & $98\%$ & $98\%$&0.035& 0.027\\
 & 50 & 0.033 & 0.033 & $98\%$ & $98\%$ &0.032&0.029\\
 & Standard & 0.022 & 0.022 & $97\%$ & $96\%$ &0.040& 0.031\\ 
\multirow{3}{*}{$\rho_{\beta}=2$} & 10 & 0.13 & 0.13 & $98\%$ & $98\%$&0.20 &0.25 \\
 & 50 & 0.13 & 0.13 & $98\%$ & $98\%$  &0.20&0.24\\
 & Standard & 0.10 & 0.13 & $98\%$ & $98\%$  &0.21 &0.25\\ 
\multirow{3}{*}{$\nu_{\beta}=0.5$} & 10 & 0.038 & 0.038 & $96\%$ & $96\%$  &0.040&0.050\\
 & 50 & 0.038 & 0.038 & $96\%$ & $98\%$ &0.044&0.052\\
 & Standard & 0.038 & 0.058 & $98\%$ & $96\%$ & 0.043&0.053\\ 
 \multicolumn{8}{l}{MAD: Mean Absolute Deviation;}\\
  \multicolumn{8}{l}{MCSD: Monte Carlo Standard Deviation;}\\
 \multicolumn{8}{l}{SWP: Spatial Wishart Process Model;}\\
 \multicolumn{8}{l}{CDP: Cholesky Decomposition Process Model.}
\end{tabular}
\end{table}

\begin{table}[t!]
\caption{Non-asymptotic ($m=3$) simulation results for spatially-varying coefficients with the data generated from the Cholesky decomposition process model or the spatial Wishart process model. The results are summarized in terms of mean absolute deviation of posterior mean estimates, $95\%$ posterior coverage, and Monte Carlo standard deviation. The values are averaged over replications, voxels ($n$), and covariates ($d$).}\label{t:sim1_non}
\centering
\begin{tabular}{cccccccc}
\multirow{2}{*}{Parameter}          & \multirow{2}{*}{$q$ } & \multicolumn{2}{c}{MAD} & \multicolumn{2}{c}{Coverage $95\%$} & \multicolumn{2}{c}{MCSD}  \\
                                    &                       & CDP   & SWP             & CDP     & SWP                       & CDP  & SWP                \\
\multirow{3}{*}{$\bm{\beta}_{11}$ } & 10                    & 0.23~ & 0.61            & $93\%$  & $100\%$                   & 0.22 & 0.75               \\
                                    & 50                    & 0.22  & 0.60            & $95\%$  & $100\%$                   & 0.21 & 0.75               \\
                                    & Standard              & 0.22  & 0.58            & $96\%$  & $100\%$                   & 0.22 & 0.78               \\
\multirow{3}{*}{$\bm{\beta}_{22}$ } & 10                    & 0.22  & 0.74            & $95\%$  & $100\%$                   & 0.22 & 0.75               \\
                                    & 50                    & 0.22  & 0.72            & $95\%$  & $100\%$                   & 0.23 & 0.75               \\
                                    & Standard              & 0.22  & 0.70            & $95\%$  & $100\%$                   & 0.22 & 0.75               \\
\multirow{3}{*}{$\bm{\beta}_{33}$ } & 10                    & 0.23  & 1.06            & $94\%$  & $100\%$                   & 0.22 & 0.75               \\
                                    & 50                    & 0.23  & 0.95            & $92\%$  & $100\%$                   & 0.23 & 0.76               \\
                                    & Standard              & 0.22  & 0.92            & $93\%$  & $100\%$                   & 0.23 & 0.75               \\
\multirow{3}{*}{$\bm{\beta}_{21}$ } & 10                    & 0.22  & 0.12            & $99\%$  & $100\%$                   & 0.24 & 0.18               \\
                                    & 50                    & 0.23  & 0.13            & $97\%$  & $100\%$                   & 0.24 & 0.15               \\
                                    & Standard              & 0.23  & 0.13            & $94\%$  & $100\%$                   & 0.23 & 0.17               \\
\multirow{3}{*}{$\bm{\beta}_{31}$ } & 10                    & 0.23  & 0.24            & $99\%$  & $100\%$                   & 0.22 & 0.17               \\
                                    & 50                    & 0.22  & 0.15            & $95\%$  & $100\%$                   & 0.23 & 0.13               \\
                                    & Standard              & 0.12  & 0.15            & $96\%$  & $100\%$                   & 0.23 & 0.17               \\
\multirow{3}{*}{$\bm{\beta}_{32}$ } & 10                    & 0.23  & 0.19            & $99\%$  & $100\%$                   & 0.22 & 0.19               \\
                                    & 50                    & 0.23  & 0.19            & $99\%$  & $100\%$                   & 0.23 & 0.18               \\
                                    & Standard              & 0.22  & 0.19            & $93\%$  & $100\%$                   & 0.24 & 0.18               \\
\multicolumn{8}{l}{MAD: Mean Absolute Deviation;}                                                                                                       \\
\multicolumn{8}{l}{MCSD: Standard Deviation;}                                                                                                           \\
\multicolumn{8}{l}{SWP: Spatial Wishart Process Model;}                                                                                                 \\
\multicolumn{8}{l}{CDP: Cholesky Decomposition Process Model.}                                                                                         
\end{tabular}
\end{table}

\begin{table}[t!]
\caption{Non-asymptotic ($m=3$) simulation results for spatial parameters with the data generated from the Cholesky decomposition process model or the spatial Wishart process model. The results are summarized in terms of mean absolute deviation of posterior mean estimates, $95\%$ posterior coverage, and Monte Carlo standard deviation. The values are averaged over replications.}\label{t:sim2_non}
\centering
\begin{tabular}{cccccccc}
\multirow{2}{*}{Parameter}          & \multirow{2}{*}{$q$ } & \multicolumn{2}{c}{MAD} & \multicolumn{2}{c}{Coverage $95\%$} & \multicolumn{2}{c}{MCSD}  \\
                                    &                       & CDP    & SWP            & CDP     & SWP                       & CDP   & SWP               \\
\multirow{3}{*}{$\rho_u=2$ }        & 10                    & 0.20~  & 0.68           & $98\%$  & $98\%$                    & 0.19  & 0.61              \\
                                    & 50                    & 0.20   & 0.66           & $96\%$  & $100\%$                   & 0.19  & 0.62              \\
                                    & Standard              & 0.14   & 0.66           & $94\%$  & $98\%$                    & 0.20  & 0.60              \\
\multirow{3}{*}{$\nu_u=0.5$ }       & 10                    & 0.033~ & 0.23           & $98\%$  & $100\%$                   & 0.027 & 0.056             \\
                                    & 50                    & 0.033  & 0.24           & $98\%$  & $100\%$                   & 0.029 & 0.055             \\
                                    & Standard              & 0.022~ & 0.22           & $96\%$  & $100\%$                   & 0.031 & 0.055             \\
\multirow{3}{*}{$\rho_{\beta}=2$ }  & 10                    & 0.13   & 0.63           & $98\%$  & $100\%$                   & 0.25  & 0.59              \\
                                    & 50                    & 0.13   & 0.53           & $93\%$  & $100\%$                   & 0.24  & 0.59              \\
                                    & Standard              & 0.10   & 0.53           & $95\%$  & $100\%$                   & 0.24  & 0.58              \\
\multirow{3}{*}{$\nu_{\beta}=0.5$ } & 10                    & 0.038~ & 0.23           & $96\%$  & $100\%$                   & 0.050 & 0.055             \\
                                    & 50                    & 0.038  & 0.24~          & $94\%$  & $100\%$                   & 0.052 & 0.054             \\
                                    & Standard              & 0.058~ & 0.23           & $96\%$  & $100\%$                   & 0.053 & 0.055             \\
\multicolumn{8}{l}{MAD: Mean Absolute Deviation;}                                                                                                       \\
\multicolumn{8}{l}{MCSD: MCStandard Deviation;}                                                                                                         \\
\multicolumn{8}{l}{SWP: Spatial Wishart Process Model;}                                                                                                 \\
\multicolumn{8}{l}{CDP: Cholesky Decomposition Process Model.}                                                                                         
\end{tabular}
\end{table}

Next, we compare the performance of the Cholesky decomposition process model and the univariate spatially-varying coefficient model \citep{gelfand2003spatial}. In clinical studies and neuroimaging, the most interesting covariate effect is the drug-use  effect ($x_{i,drug}$) \citep{brick1998drugs}. The six coefficients comprehensively but not concisely describe the local covariate effects, which may not be affirmative to clinicians who prefer scalar quantities (e.g., fractional anisotropy). However, since the six coefficients capture the covariate effects without information loss, our method can accurately project the information onto any clinically meaningful scalar quantity. One of the useful quantities is fractional anisotropy, projecting a positive definite matrix onto $[0,1]$, defined as 
\begin{equation}
{\displaystyle {f_{FA}(\bm{A}})={\sqrt {\frac {1}{2}}}{\frac {\sqrt {(\lambda _{1}-\lambda _{2})^{2}+(\lambda _{2}-\lambda _{3})^{2}+(\lambda _{3}-\lambda _{1})^{2}}}{\sqrt {\lambda _{1}^{2}+\lambda _{2}^{2}+\lambda _{3}^{2}}}}},
\end{equation}
where $\{\lambda_1,\lambda_2,\lambda_3\}$ are the eigenvalues of a diffusion tensor $\bm{A}$ \citep{ennis2006orthogonal}. To demonstrate this, we estimate the treatment effect of cocaine use on each voxel $\bm{s}$ in terms of fractional anisotropy, denoted as $\delta_{FA}(\bm{s})=f_{FA}(\bm{\Sigma}^{(1)}({\bm{s}}))-f_{FA}(\bm{\Sigma}^{(0)}({\bm{s}}))$. Assuming $\bm{X}_i'$ as the covariates excluding drug use, the term $\bm{\Sigma}^{(d)}({\bm{s}})=\frac{1}{N}\sum_{i=1}^N\mathbb{E}[\bm{\Sigma}(\bm{s})\mid \bm{X}_i',x_{drug}=d]$ is to describe the averaged (over subjects) mean matrix at voxel $\bm{s}$ under drug-use  status $d\in\{0,1\}$. We use a Markov chain Monte Carlo-based outcome regression estimator \citep{rotnitzky1998semiparametric} to estimate $\delta_{FA}(\bm{s})$, defined as 
\begin{equation}
\hat{\delta}_{FA}(\bm{s})=\frac{1}{N}\sum_{i=1}^{N}\left(\mathbb{E}[f_{FA}[\bm{\Sigma}_i(\bm{s})]\mid \bm{x}_{i,drug}=1,rest]-\mathbb{E}[f_{FA}[\bm{\Sigma}_i(\bm{s})]\mid \bm{x}_{i,drug}=0,rest]\right),
\end{equation}
where the expectation can be empirically obtained by Markov chain Monte Carlo samples.

\begin{figure}[t!]
    \centering
    \begin{subfigure}[t!]{0.45\textwidth}
       \includegraphics[width=\textwidth]{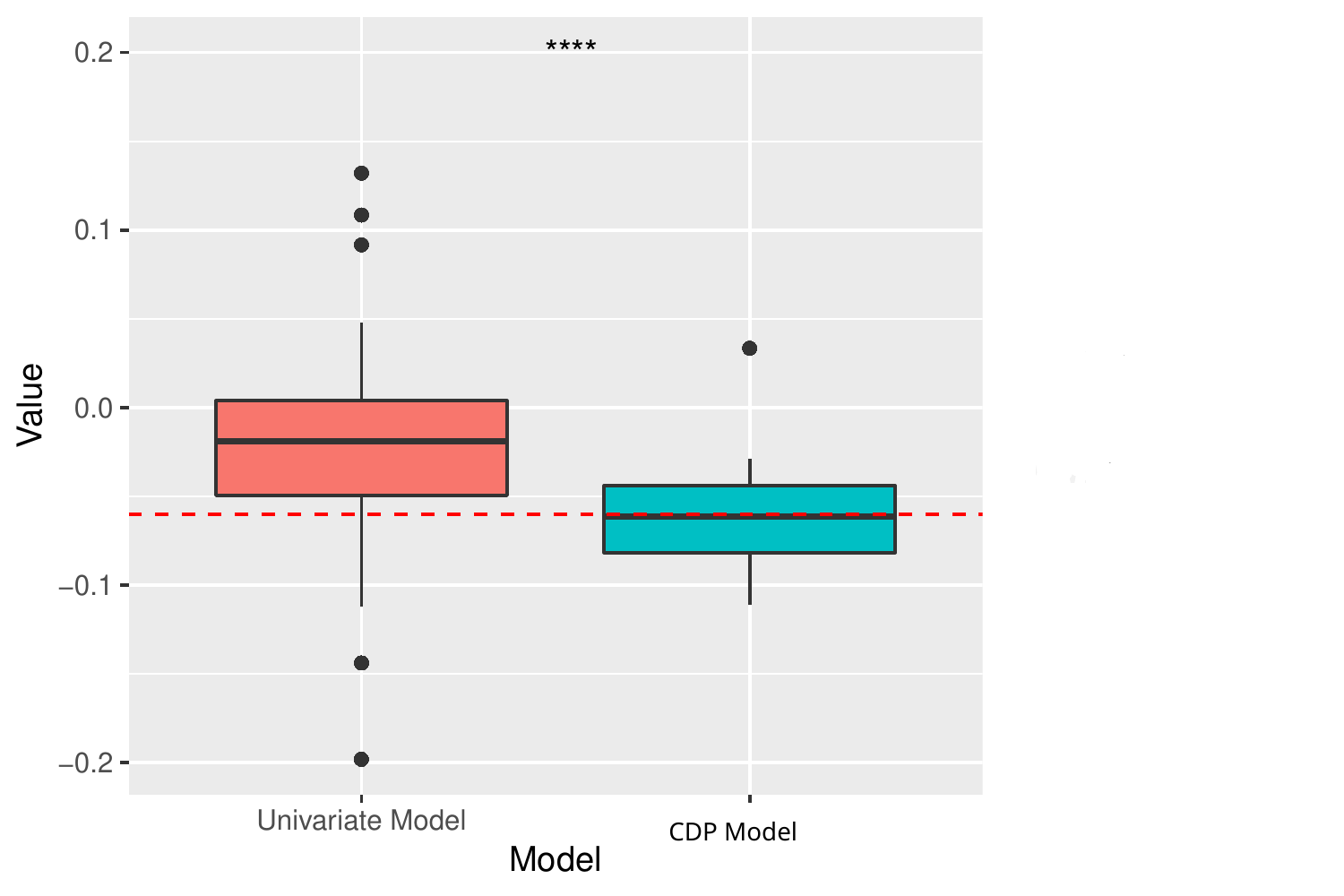}
          \caption{$\delta_{FA}(\bm{s})$ for $\bm{s}\in\mathcal{S}$.}
    \end{subfigure}
     \begin{subfigure}[t!]{0.45\textwidth}
    \includegraphics[width=\textwidth]{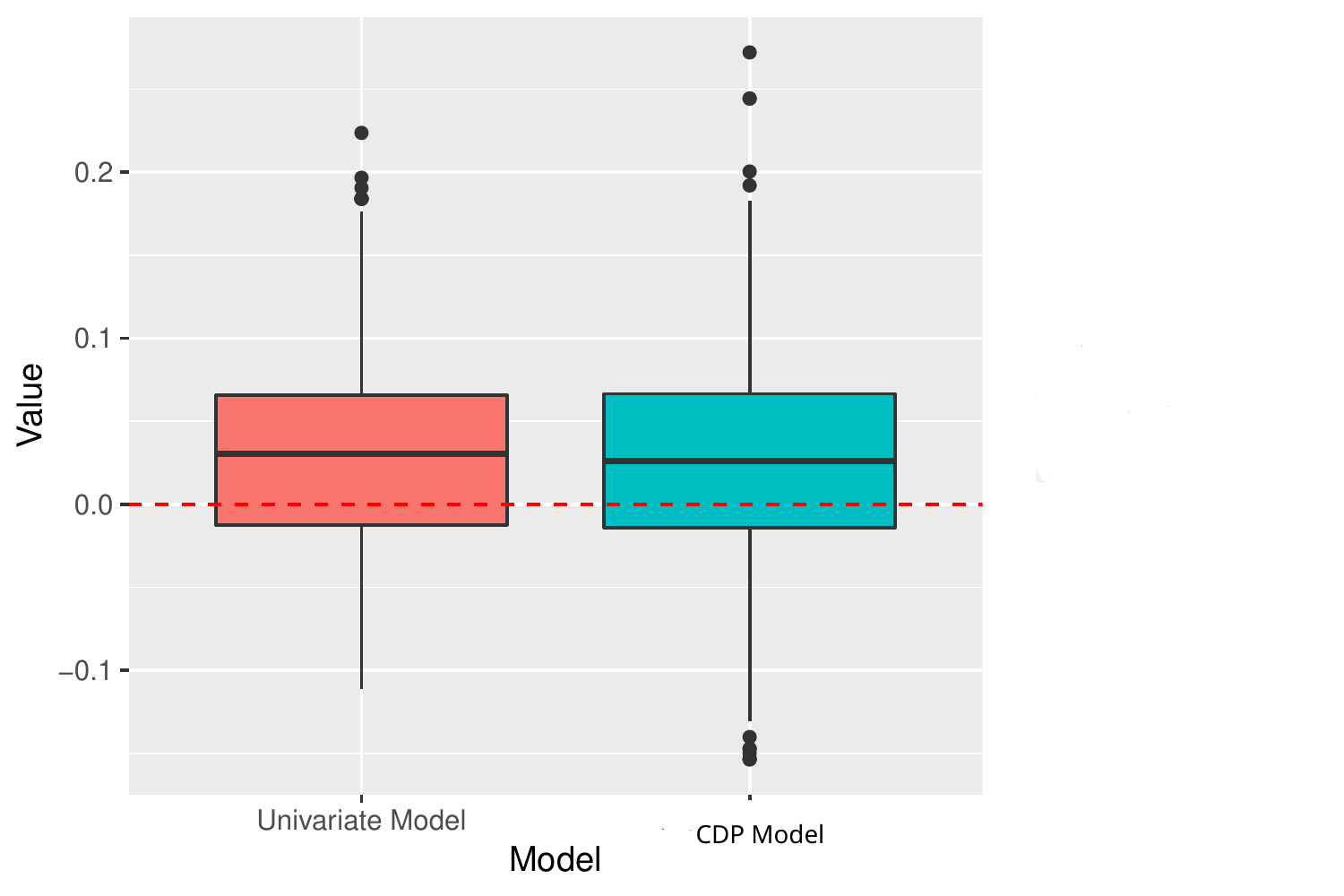}
    \caption{$\delta_{FA}(\bm{s})$ for $\bm{s}\notin\mathcal{S}$}
    \end{subfigure}
    \caption{Estimates of $\delta_{FA}(\bm{s})$ produces by the Cholesky decomposition process model and the univariate model. The red dashed lines are the true values. $\mathcal{S}$ is a set of spatial locations inside a $4 \times 4$ region in the middle of the image.}\label{fig:dFA}
\end{figure}

Since the spatial matrix-variate methods in terms of coefficients estimation have consistent results, we simply use the results of the Cholesky decomposition process model with $q=10$ and $m=50$ for a concise illustration. We plot the posterior means of $\delta_{FA}(\bm{s})$ in Figure \ref{fig:dFA}, combining all voxels $\bm{s}$ and replications. We compare it to the univariate spatially-varying coefficient model \citep{gelfand2003spatial} with logit transformation of fractional anisotropy as responses and its associated Markov chain Monte Carlo-based outcome regression estimator \citep{rotnitzky1998semiparametric} is \begin{equation}
    \hat{\delta}_{FA}(\bm{s})=\frac{1}{N}\sum_{i=1}^{N}\left(\mathbb{E}[f_{logit}^{-1}[y_i(\bm{s})]\mid \bm{x}_{i,drug}=1,rest]-\mathbb{E}[f_{logit}^{-1}[y_i(\bm{s})]\mid \bm{x}_{i,drug}=0,rest]\right),
    \end{equation}
    where $y_i(\bm{s})$ is response and $f_{logit}$ is the logit transformation. In Figure \ref{fig:dFA}, the Cholesky decomposition process model produces more precise estimates for $\delta_{FA}(\bm{s})$ for $\bm{s}\in\mathcal{S}$ with smaller uncertainties, revealing that utilizing the whole matrix information plays a key role in detecting covariate effects. This claim is further verified in real data analysis (see Figure \ref{fig:FA}).

\section{Application to the Cocaine User Data}
\label{s:app}
In this section, we apply the model to a data set of cocaine users \citep{ma2017preliminary}. The data are provided by the Institute for Drug and Alcohol Studies of Virginia Commonwealth University (VCU). Eleven cocaine users and eleven non-cocaine users participated in this study. Besides their cocaine-use status, their age and education years are also recorded. Following the conventions in cocaine use studies \citep[e.g.,][]{lane2010diffusion,ma2017preliminary}, we focus on the corpus callosum, a brain region playing important roles such as transferring motor, sensory, and cognitive information between the brain hemispheres. This region contains $15,273$ voxels.

We first fit the data to the Cholesky decomposition process model to investigate the covariate effects and spatial dependence among voxels. We set the design matrix $\bm{X}_i$ as $[1,x_{i,drug}, x_{i,age}, x_{i,edu}, x_{i,handedness},x_{gender}]$, representing the intercept, drug-use  ($x_{i,drug}=1$ if subject $i$ is a cocaine user, otherwise $x_{i,drug}=0$), age, education years, handedness ($x_{i,handedness}=1$ if subject $i$ is a left-handed, otherwise $x_{i,handedness}=0$), and gender ($x_{i,gender}=1$ if subject $i$ is female, otherwise $x_{i,gender}=0$). We set $q=50$ for Vecchia's approximation and totally $8,000$ Markov chain Monte Carlo samples are collected after $3,000$ samples as burn-in.

\begin{figure}[t!]
    \centering
    \begin{subfigure}[t!]{0.24\textwidth}
       \includegraphics[width=\textwidth]{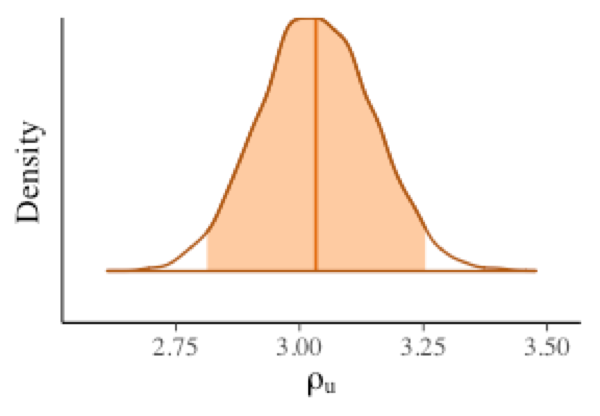}
          \caption{$\rho_u$.}\label{fig:MCMC_rho_u}
    \end{subfigure}
     \begin{subfigure}[t!]{0.24\textwidth}
    \includegraphics[width=\textwidth]{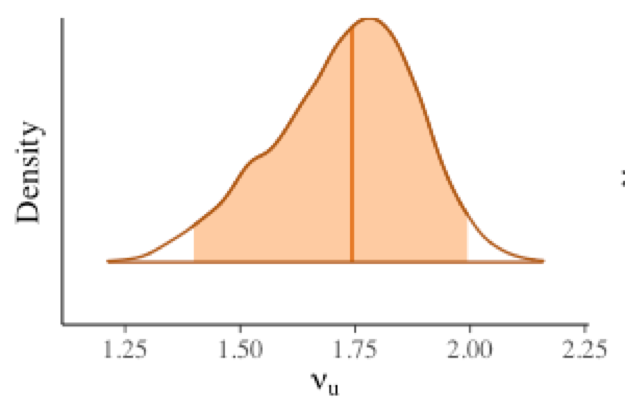}
    \caption{$\nu_u$.}\label{fig:MCMC_smooth_u}
    \end{subfigure}
        \begin{subfigure}[t!]{0.24\textwidth}
       \includegraphics[width=\textwidth]{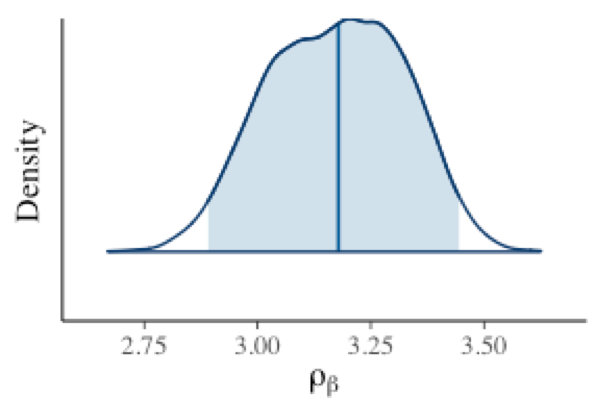}
          \caption{$\rho_\beta$.}\label{fig:MCMC_rho_beta}
    \end{subfigure}
     \begin{subfigure}[t!]{0.24\textwidth}
    \includegraphics[width=\textwidth]{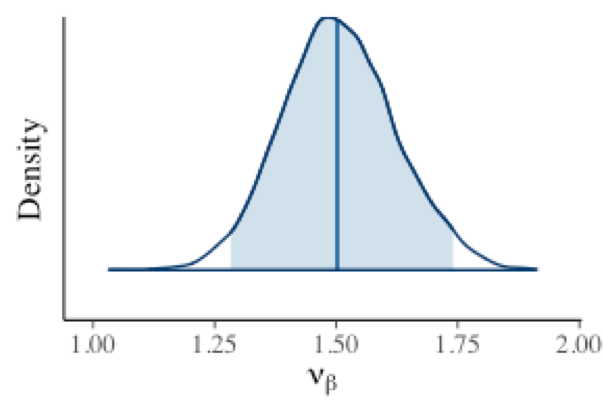}
    \caption{$\nu_\beta$.}\label{fig:MCMC_smooth_beta}
    \end{subfigure}
    
    \caption{The posterior densities of spatial parameters for the cocaine user data.}\label{fig:MCMC}
\end{figure}

To understand the spatial dependence of the DTs, we plot the posterior density of the spatial dependence parameters $\rho_u$, $\nu_u$, $\rho_\beta$, and $\nu_\beta$ in Figure \ref{fig:MCMC}, where the posterior mean estimates are $3.034$, $1.75$, $3.17$, $1.51$, respectively, and the $95\%$ credible regions are $[2.81,3.19]$, $[1.39,1.93]$, $[2.89,3.38]$, $[1.28,1.67]$, respectively. The result reveals that the spatial dependence of the residual process and mean process are strong and smooth. The approximation of the Cholesky decomposition process to the spatial Wishart process holds if $m=\sigma_m^{-2}$ is large. For our data, the posterior $95\%$ interval of $m=\sigma_m^{-2}$ is $[33.13,33.89]$, supporting the asymptotic approximation.

\begin{figure}[t!]
    \centering
     \begin{subfigure}[t!]{0.6\textwidth}
    \includegraphics[width=\textwidth]{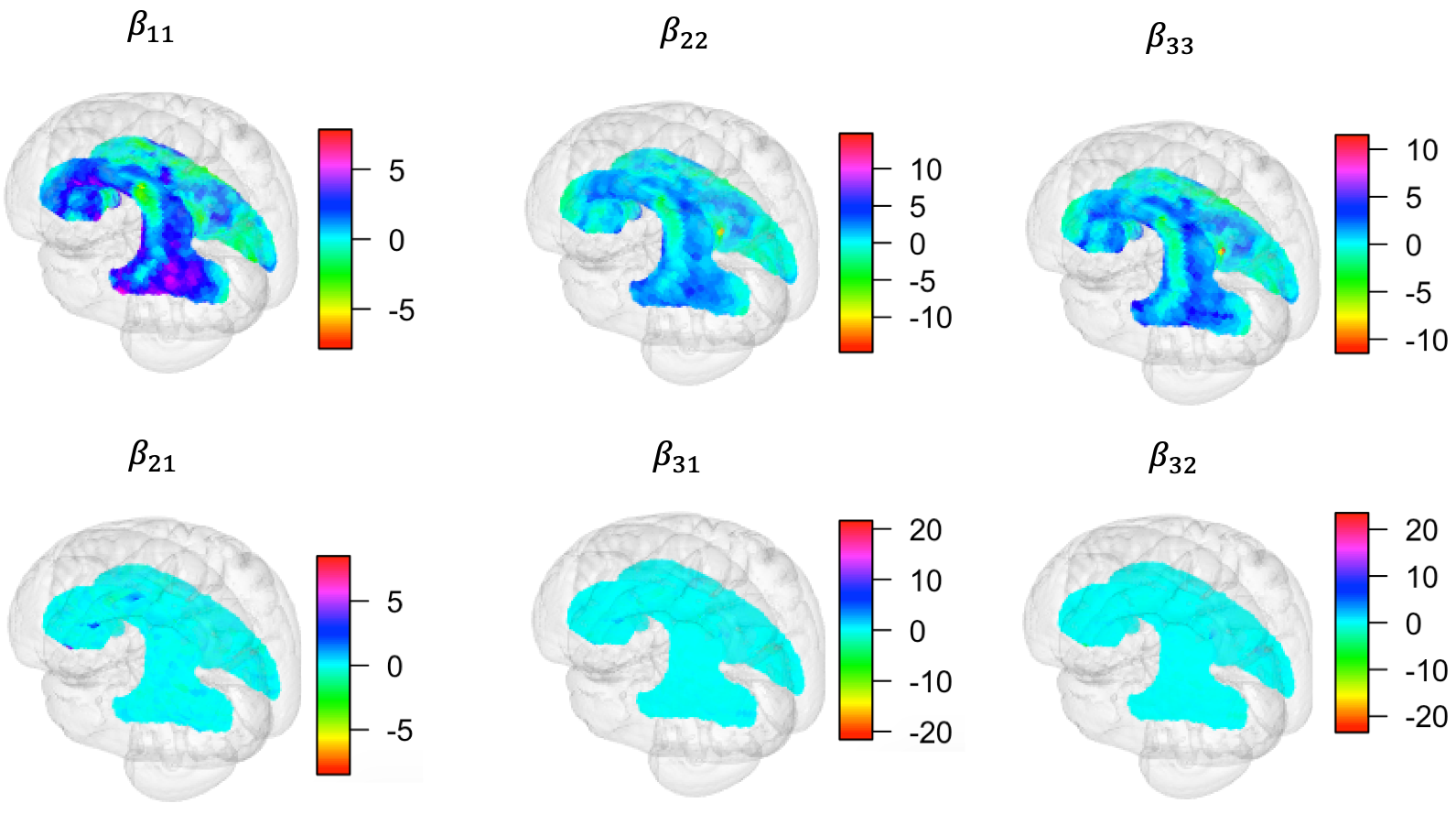}
    \caption{The covariate effects of age.}\label{fig:age}
    \end{subfigure}
     ~
        \begin{subfigure}[t!]{0.6\textwidth}
       \includegraphics[width=\textwidth]{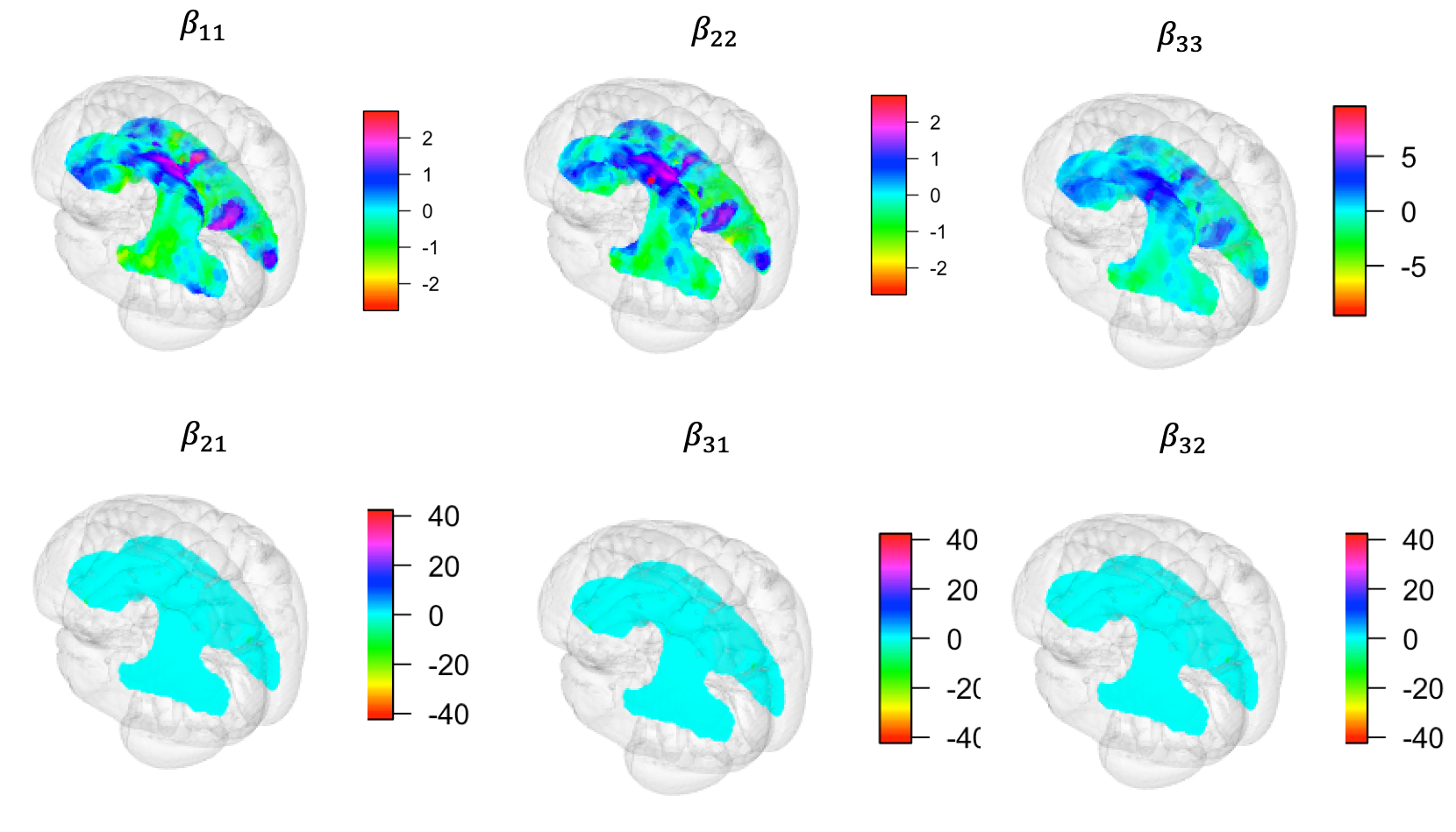}
          \caption{The covariate effects of gender.}\label{fig:gender}
    \end{subfigure}
    
    \caption{The covariate effects of biological attributes on DT expressing by the posterior z-scores of six spatially-varying coefficients.}\label{fig:covariate1}
\end{figure}

\begin{figure}
    \centering
       \begin{subfigure}[t!]{0.6\textwidth}
       \includegraphics[width=\textwidth]{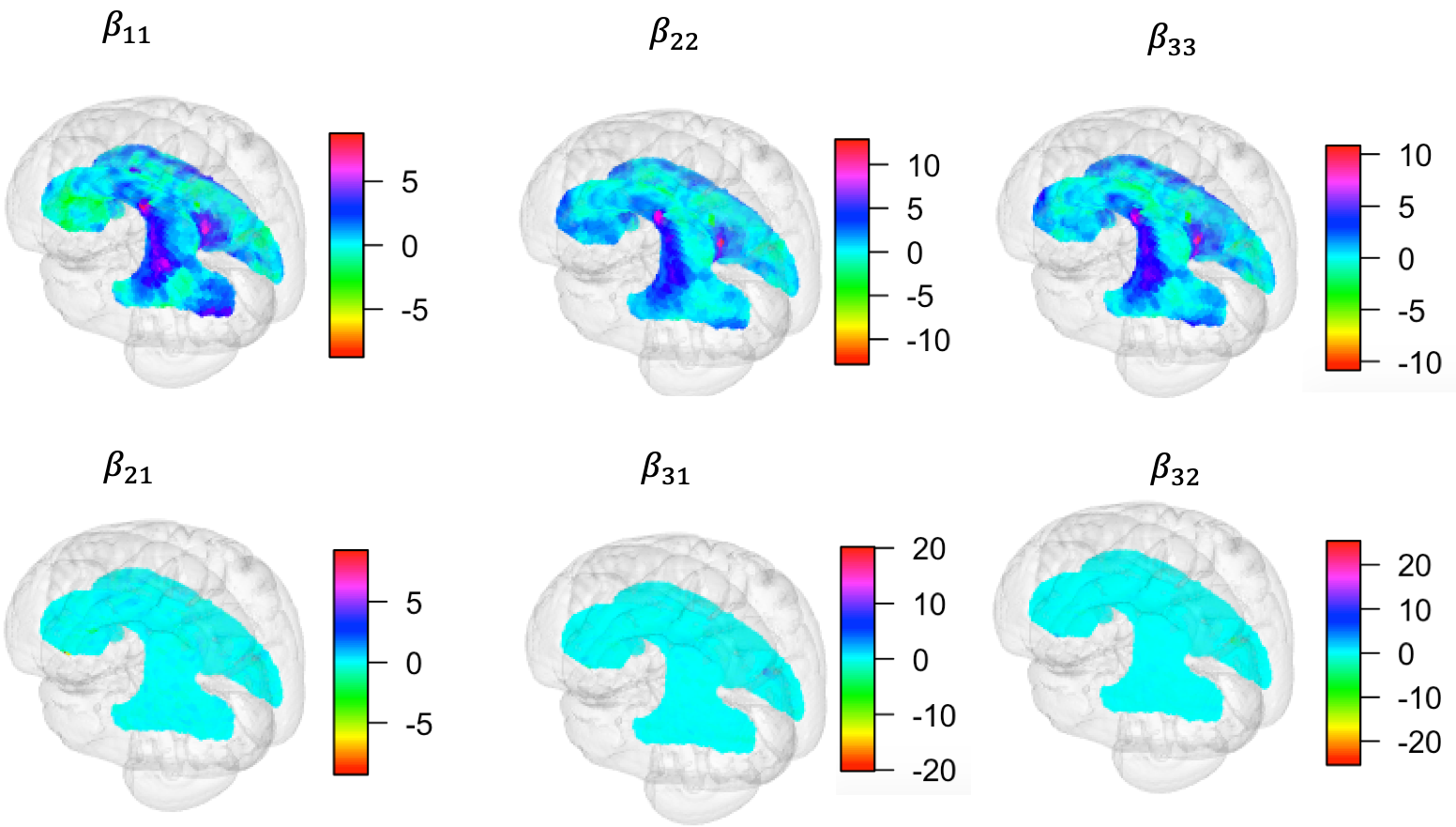}
          \caption{The covariate effects of cocaine use.}\label{fig:cocaine}
    \end{subfigure}
    ~
        \begin{subfigure}[t!]{0.6\textwidth}
       \includegraphics[width=\textwidth]{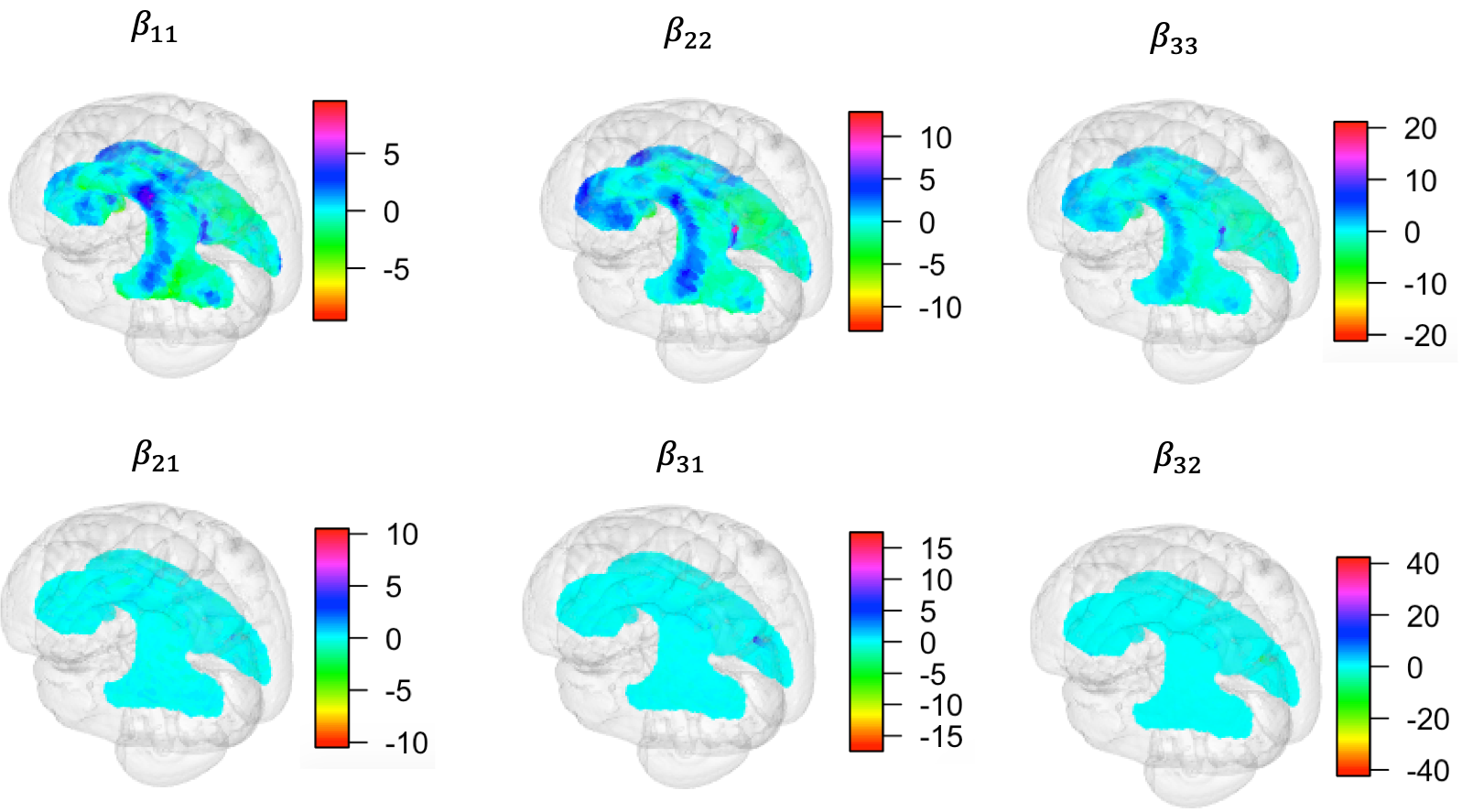}
          \caption{The covariate effects of education years.}\label{fig:education}
    \end{subfigure}
  ~
        \begin{subfigure}[t!]{0.6\textwidth}
       \includegraphics[width=\textwidth]{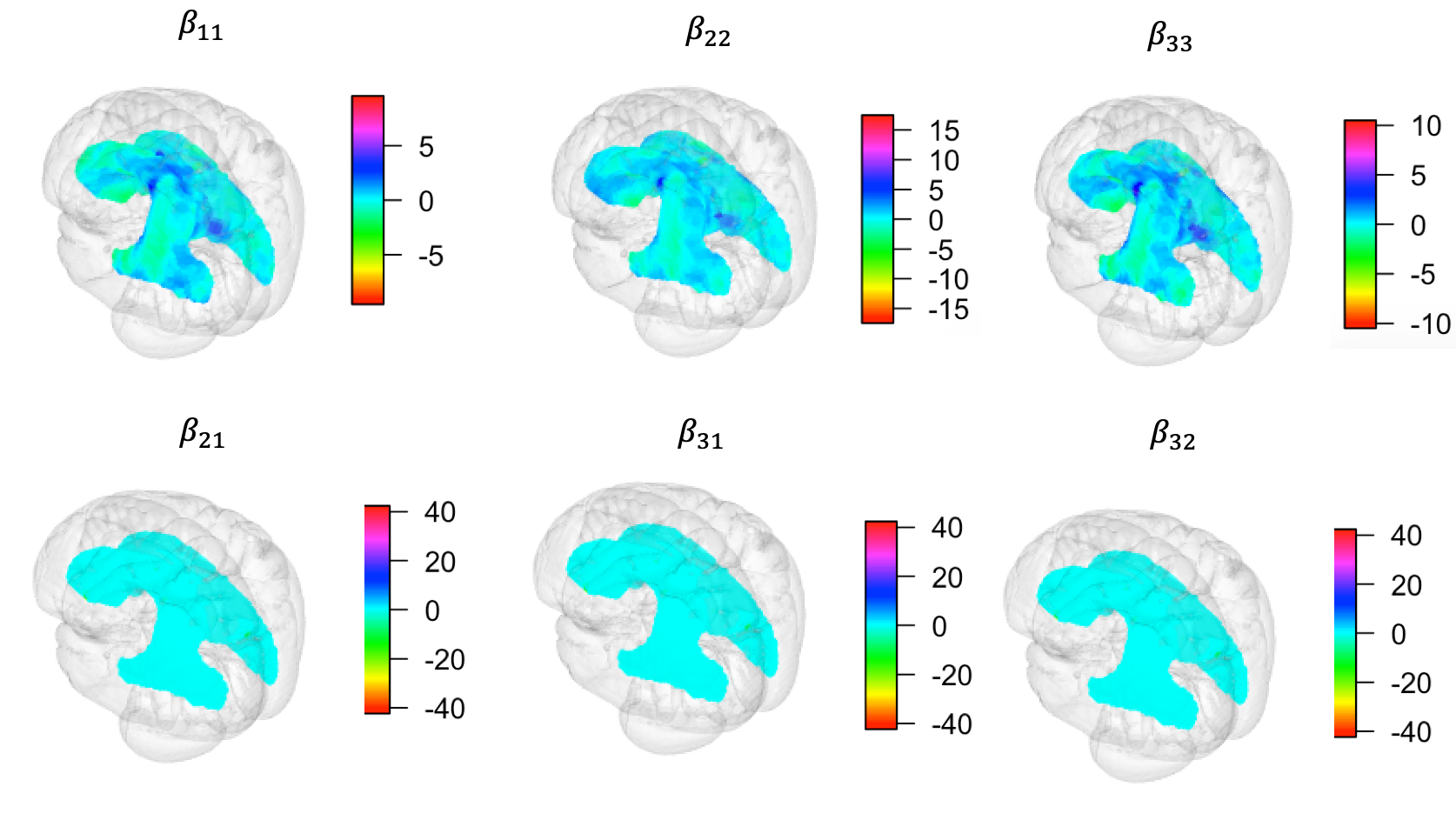}
          \caption{The covariate effects of handedness.}\label{fig:hand}
    \end{subfigure}
   \caption{The covariate effects of social attributes on DT expressing by the posterior z-scores of six spatially-varying coefficients.}\label{fig:covariate2}
\end{figure}

To compare the Bayesian mean estimates to 0 and quantify their uncertainties, the covariate effects on DTs expressed by their posterior z-scores\footnote{$\frac{\mathbb{E}[\theta\mid .]}{SD[\theta\mid .]}$ where $\mathbb{E}[\theta\mid .]$ is the posterior mean and $SD[\theta\mid .]$ is the posterior standard deviation.} \citep{louis1984estimating} are visualized in Figure \ref{fig:covariate1} and \ref{fig:covariate2}, which are smooth over voxels. Among these covariates, cocaine use is the covariate with the most significant impact, where the diagonal coefficients have many absolutely large posterior z-scores located at some regions. Education years has no effect in most areas but a powerful impact at certain areas (see $\beta_{11}$ and $\beta_{31}$), which needs a further scientific investigation. Unlike education years, the effect of age has a significant impact all over the corpus callosum, which may indicate the effect of age on brain structure is little but covers the whole area. Gender and handedness have comparatively small overall effects in comparison to the others but have strong effects in certain areas.

\begin{figure}[t!]
    \centering
    \begin{subfigure}[t!]{0.45\textwidth}
       \includegraphics[width=\textwidth]{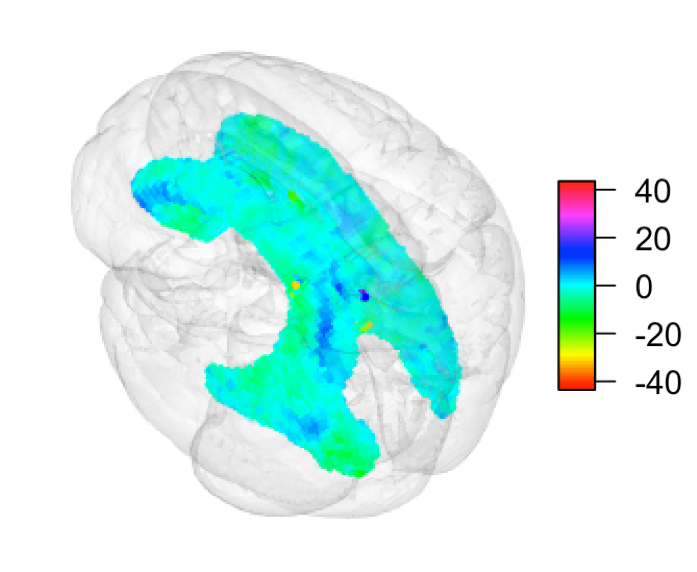}
          \caption{The posterior z-scores of $\delta_{FA}$ based on the Cholesky decomposition process model.}\label{fig:FA_working}
    \end{subfigure}
    ~
     \begin{subfigure}[t!]{0.45\textwidth}
    \includegraphics[width=\textwidth]{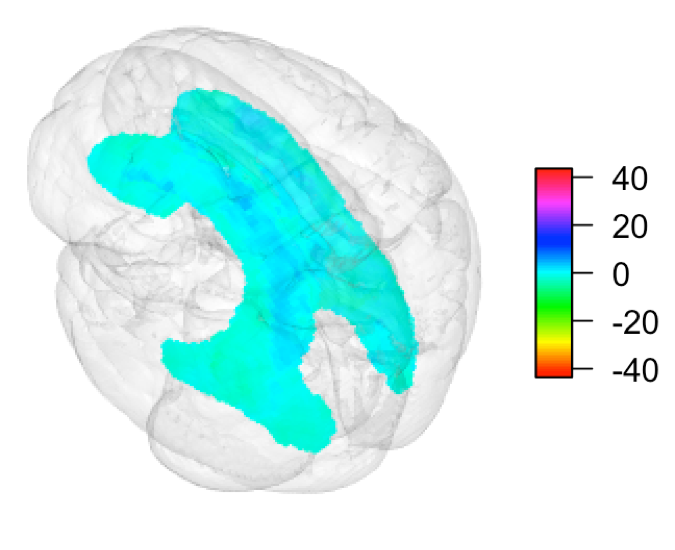}
    \caption{The posterior z-scores of $\delta_{FA}$ based on the univariate model.}\label{fig:FA_uni}
    \end{subfigure}
   
    \caption{The posterior z-scores of $\delta_{FA}$.}\label{fig:FA}
\end{figure}

Furthermore, we use the Markov chain Monte Carlo-based outcome regression estimator to estimate the drug-use effect $\delta_{FA}(\bm{s})$. Figure \ref{fig:FA_working} provides the posterior z-scores of $\hat{\delta}_{FA}(\bm{s})$. In comparison, Figure \ref{fig:FA_uni} provides the posterior z-scores of the Bayesian outcome regression estimator of $\delta_{FA}(\bm{s})$ based on the univariate spatially-varying coefficients model \citep{gelfand2003spatial} with the logit transformation of the DTs' fractional anisotropy as responses. The Cholesky decomposition process model provides a definite region where cocaine use has a significant effect with strong intensity, whereas the effect detected by the univariate model is less distinct. This demonstrates the advantages of the matrix-variate modeling over univariate modeling. The regions of differences are located at the splenium, a component at the posterior end of the corpus callosum, indicating group differences between cocaine users and non-cocaine users. This result is also consistent with previous clinical studies on cocaine use \citep[e.g.,][]{lane2010diffusion}, validating the clinical reliability of our proposal.

\section{Discussion}
\label{s:diss}
In this paper, we propose geostatistical modeling for positive definite matrices, with applications to DTI. Considering that the literature of spatial modeling of positive definite matrices is sparse, the spatial Wishart process as a random field for spatially dependent positive definite matrices offers a useful and elegant approach. We further propose a Cholesky decomposition process model whose responses are the Cholesky decomposition of positive definite matrices, overcoming the problematic issues caused by the intractable probability density function of the spatial Wishart process. Both the simulation studies and real data application demonstrate the effectiveness in spatial Bayesian inference of positive definite matrices. 

Besides neuroimaging application, our work also makes a theoretical contribution to positive definite matrix-variate modeling. A bottleneck of the spatial Wishart process model is that the probability density function is intractable, which here is resolved by proposing the Cholesky decomposition process model. Other attempts are mostly to investigate the special cases where the latent covariance matrix has a certain form \citep[e.g.,][]{mathai1991multivariate,furman2008multivariate}, compromising the flexibility. Meanwhile, \citet{yu2004empirical} are proposing characteristic function-based parameter estimation approaches for models whose probability density function is intractable but characteristic function is elegant. In Appendix I, we give the characteristic function of the spatial Wishart process which is simple, providing an alternative inference approach. 

The most important contribution is that we have shown that the Cholesky decomposition process model and the spatial Wishart process model are asymptotically equivalent. This might provide an insight that some positive definite matrix-variate models \citep[e.g.,][]{karagiannidis2003efficient,smith2007distribution,kuo2007joint} can be approximated to Gaussian processes, allowing statistical and computational benefits brought from Gaussian processes.

\section*{Supplementary material}
\label{SM}

\appendix

Before showing the proofs, we emphasize again that the Wishart distribution in this paper is a parameterized Wishart distribution in terms of mean matrix and degrees of freedom. That is, let $\bm{A}\sim\mathcal{W}_p(\bm{V},n)$, the PDF is 
$${\displaystyle f(\mathbf {A} )={\frac {\mid \mathbf {A} \mid ^{(n-p-1)/2}e^{-\operatorname {tr} ([\mathbf {V}/n]^{-1}\mathbf {A} )/2}}{2^{\frac {np}{2}}\mid {\mathbf {V}/n }\mid ^{n/2}\Gamma _{p}({\frac {n}{2}})}}},$$
with $\mathbb{E}\bm{A}=\bm{V}$. This parameterization is convenient in our model. But this is different from the classic Wishart distribution specified in most classic textbooks of multivariate statistics \citep[i.e.,][]{mardia1980multivariate,anderson1984introduction,gupta1999matrix,eaton2008multivariate}, where the mean is $n\bm{V}$. However, there is no difficulty in expressing the results of classic Wishart distribution using this parameterization of the Wishart distribution.

\section*{Appendix I}
\subsection*{Properties of spatial Wishart process}
Before giving the proof, we give the characteristic function of $[\bm{U}(\bm{s}_1), ..., \bm{U}(\bm{s}_n)]$ (Corollary \ref{theorem:char}), making the proof more convenient.

\begin{corollary}
    \label{theorem:char}
       The characteristic function of $[\bm{U}(\bm{s}_1), ..., \bm{U}(\bm{s}_n)]$ is
\begin{equation}
\label{eq:chara}
    \phi(\bm{T}_{\bm{s}_1}, ..., \bm{T}_{\bm{s}_n})=\mathbb{E}\exp tr[\imath(\bm{T}_{\bm{s}_1}\bm{U}(\bm{s}_1)+...+\bm{T}_{\bm{s}_n}\bm{U}(\bm{s}_n))]=\mid \bm{I}_{np}-2\imath\bm{T}(\bm{R}\otimes\bm{\Sigma}/m)\mid ^{-m/2},
\end{equation}
where $\imath$ is the imaginary number with $\imath^2=-1$, $\bm{T}_{\bm{s}}$ is a symmetric matrix whose the diagonal entries are $t_{ii}(\bm{s})$ and off-diagonal elements are $\frac{1}{2}t_{ij}(\bm{s})$, $\bm{T}$ is a block diagonal matrix composed of $[\bm{T}_{\bm{s}_1}, ..., \bm{T}_{\bm{s}_n}]$, and $\bm{R}$ is the spatial correlation matrix of locations $\{\bm{s}_1, ...,  \bm{s}_n\}$ constructed by $\mathcal{K}(\bm{s},\bm{s}'\mid \bm{\Phi})$.
\end{corollary}

\begin{proof}

As an extension of \citet[][Equation 2]{krishnaiah1961remarks}, the proof largely relies on \citet[][Section 7.3.1]{anderson1984introduction}, which provides the derivation of characteristic function of a Wishart distribution. We extend Equation (4) of \citet[][Section 7.3.1]{anderson1984introduction} to a spatial case:
\begin{equation}
\label{eq:ch}
\begin{aligned}
\phi(\bm{T}_{\bm{s}_1}, ..., \bm{T}_{\bm{s}_n})&=\prod_{j=1}^m\mathbb{E}\exp tr[\imath(\bm{Z}_j(\bm{s}_1)^T\bm{T}_{\bm{s}_1}\bm{Z}_j(\bm{s}_1)/m+...+\bm{Z}_j(\bm{s}_n)^T\bm{T}_{\bm{s}_n}\bm{Z}_j(\bm{s}_n)/m)]  \\
&=(\mathbb{E}\exp tr[\imath(\bm{Z}(\bm{s}_1)^T\bm{T}_{\bm{s}_1}\bm{Z}(\bm{s}_1)/m+...+\bm{Z}(\bm{s}_n)^T\bm{T}_{\bm{s}_n}\bm{Z}(\bm{s}_n)/m)])^m  \\
&= (\mathbb{E}\exp tr[\imath\bm{Z}^T\bm{T}\bm{Z}/m])^m\\
&= (\mathbb{E}\exp tr[\imath\tilde{\bm{Z}}^T\bm{T}\tilde{\bm{Z}}])^m,
\end{aligned}
\end{equation}
where $\tilde{\bm{Z}}$ is a stack vector such as $\tilde{\bm{Z}}=[\bm{Z}(\bm{s}_1)^T/m^{1/2}, ..., \bm{Z}(\bm{s}_n)^T/m^{1/2}]^T$ following a mean-zero Gaussian distribution whose covariance matrix is $\bm{R}\otimes\bm{\Sigma}/m$. Then we can continue to work on the derivation following \citet[][Section 7.3.1]{anderson1984introduction} and finally get to Equation (11) of \citet[][Section 7.3.1]{anderson1984introduction} under the spatial case.
\end{proof}

\begin{proof}[of Theorem~\ref{thm:swp}]

    There are two conditions (K1 and K2) in Kolmogorov's extension theorem:
    \begin{description}
    \item[K1:] 
     Let $\bm{s}_1, ..., \bm{s}_K\in\mathcal{D}$. We want to show for every permutation of observation indices $\pi_1, ..., \pi_K$, we have $$p(\bm{U}(\bm{s}_1), ..., \bm{U}(\bm{s}_K))=p(\bm{U}(\bm{s}_{\pi_1}), ..., \bm{U}(\bm{s}_{\pi_K})).$$
   It is equivalent to showing that $\phi(\bm{T}_{\bm{s}_1}, ..., \bm{T}_{\bm{s}_K})=\phi(\bm{T}_{\bm{s}_{\pi_1}}, ..., \bm{T}_{\bm{s}_{\pi_K}})$; Since the determinant in the characteristic function (\ref{eq:chara}) is invariant to permutation, K1 condition holds.
    \item[K2:]
   The second condition need to verify that, for every location $\bm{s}_0\in\mathcal{D}$, we have $$p(\bm{U}(\bm{s}_1), ..., \bm{U}(\bm{s}_K))=\int p(\bm{U}(\bm{s}_0),\bm{U}(\bm{s}_1), ..., \bm{U}(\bm{s}_K))d\bm{U}(\bm{s}_0).$$
        The characteristic function of $[\bm{U}(\bm{s}_0),\bm{U}(\bm{s}_1), ..., \bm{U}(\bm{s}_K))]$ is
        $\phi(\bm{T}_{\bm{s}_0},\bm{T}_{\bm{s}_1}, ..., \bm{T}_{\bm{s}_K})$.
       It is equivalent to showing that $\phi(\bm{T}_{\bm{s}_0}=\bm{0}, \bm{T}_{\bm{s}_1}, ..., \bm{T}_{\bm{s}_K})=\phi(\bm{T}_{\bm{s}_1}, ..., \bm{T}_{\bm{s}_K})$, and this follows from the form of the characteristic function (\ref{eq:chara}).
    \end{description}
    In fact, giving Equation (\ref{eq:ch}) in terms of $\bm{Z}_j$ is sufficient to show \textbf{K1} and \textbf{K2}. Giving Corollary  \ref{theorem:char} is additional.
\end{proof}

\begin{proof}[of Property \ref{prop:conti}]

    Since \citet{kent1989continuity} proved that the spatial Gaussian process $\{\bm{Z}_j(\bm{s}):\bm{s}\in \mathcal{D}\}$ holds almost-sure continuity under the condition described in the property, we have $\bm{Z}_j(\bm{s})\bm{Z}_j(\bm{s})^T$ converges almost-surely to $\bm{Z}_j(\bm{s}_0)\bm{Z}_j(\bm{s}_0)^T$ if $\mid \mid \bm{s}-\bm{s}_0\mid \mid \rightarrow 0$ given the continuous mapping theorem \citep[Theorem 2.3]{van2000asymptotic}. Given the continuous mapping theorem \citep[Theorem 2.3]{van2000asymptotic} (we know $[\bm{Z}_1(\bm{s})\bm{Z}_1(\bm{s})^T, ..., \bm{Z}_m(\bm{s})\bm{Z}_m(\bm{s})^T]^T$ converges almost-surely to $[\bm{Z}_1(\bm{s}_0)\bm{Z}_1(\bm{s}_0)^T, ..., \bm{Z}_m(\bm{s}_0)\bm{Z}_m(\bm{s}_0)^T]^T$ if $\mid \mid \bm{s}-\bm{s}_0\mid \mid \rightarrow 0$), we have $\sum_j\bm{Z}_j(\bm{s})\bm{Z}_j(\bm{s})^T/m$  converges almost-surely to $\sum_j\bm{Z}_j(\bm{s}_0)\bm{Z}_j(\bm{s}_0)^T/m$ if $\mid \mid \bm{s}-\bm{s}_0\mid \mid \rightarrow 0$, implying that $\{\bm{U}(\bm{s}):\bm{s}\in\mathcal{D}\}$ holds almost-sure continuity.
\end{proof}

\section*{Appendix II}
\subsection*{Preliminary Results}
To obtain the asymptotic results, we first establish non-asymptotic results. In particular, we give the marginal distribution (for one location) of spatial Wishart process model by giving Corollary \ref{theorem:chol2} and give the joint distribution (for multiple locations) of spatial Wishart process model by giving Theorem \ref{theorem:chol3}.

First, the results of \citet[][Lemma 8.10 and Proposition 8.11]{eaton2008multivariate} can be summarized as 

\begin{theorem}
    \label{theorem:chol}
    Suppose $\bm{A}\sim \mathcal{W}(\bm{\Sigma},m)$ and $\bm{T}$ is $\bm{A}$'s Cholesky matrix so that $\bm{T}\bm{T}^T=\bm{A}$ and $\{t_{kl}\}$ are the elements of $\bm{T}$, then
        \begin{enumerate}
            \item \textbf{Diagonal}: $t_{kk}^2\sim\mathcal{GA}(\frac{m-(k-1)}{2},2\sigma_{kk}^2/m)$, independent over $k$, where $\sigma_{kk}^2$ is the $k$-th diagonal element of $\bm{\Sigma}$;\\
            \item \textbf{Off-Diagonal}: if we partition $\bm{A}$, $\bm{T}$ and $\bm{\Sigma}$ as
            \begin{equation}
            \bm{A}=\begin{bmatrix}
            \bm{A}_{11} & \bm{A}_{12}\\
            \bm{A}_{21} & \bm{A}_{22}
            \end{bmatrix},\bm{T}=\begin{bmatrix}
            \bm{T}_{11} & \bm{0}\\
            \bm{T}_{21} & \bm{T}_{22}
            \end{bmatrix},\bm{\Sigma}=\begin{bmatrix}
            \bm{\Sigma}_{11} & \bm{\Sigma}_{12}\\
            \bm{\Sigma}_{21} & \bm{\Sigma}_{22}
            \end{bmatrix}
            \end{equation}
            where $\bm{A}_{kk}$, $\bm{T}_{kk}$, and $\bm{\Sigma}_{kk}$ are $p_k\times p_k$ square matrices, then 
            $$\bm{T}_{21}\mid \bm{T}_{11}\sim\mathcal{MN}_{p_2, p_1}(\bm{T}_{11}\bm{\Sigma}_{11}^{-1}\bm{\Sigma}_{12}, [\bm{\Sigma}_{22}-\bm{\Sigma}_{21}\bm{\Sigma}_{11}^{-1}\bm{\Sigma}_{12}]/m,\bm{I}_{p_1}).$$
        \end{enumerate}
\end{theorem}
In Theorem \ref{theorem:chol}, $\mathcal{GA}(\alpha,\beta)$ is the gamma distribution with shape parameter $\alpha$ and scale parameter $\beta$; $\mathcal{MN}_{p_r, p_c}(\bm{\mu},\bm{\Sigma}_r,\bm{\Sigma}_c)$ is the $p_r\times p_c$ matrix-variate normal distribution \citep{dawid1981some} with mean matrix $\bm{\mu}$, row covariance matrix $\bm{\Sigma}_r$ and column covariance matrix $\bm{\Sigma}_c$. To apply this result to the spatial Wishart process, let $\bm{T}_i(\bm{s})$ be the Cholesky decomposition of $\bm{A}_i(\bm{s})$ with $\bm{A}_i(\bm{s})=\bm{T}_i(\bm{s})\bm{T}_i(\bm{s})^T$ and $t_{ikl}(\bm{s})$ be the $(k,l)$-th element of $\bm{T}_i(\bm{s})$. Since $\bm{A}_i(\bm{s})$ can also be decomposed as $\bm{L}_i(\bm{s})\bm{U}_i(\bm{s})\bm{L}_i(\bm{s})^T$, we give Corollary \ref{theorem:chol2}:

\begin{corollary}
    \label{theorem:chol2}
    Suppose $\bm{A}\sim \mathcal{W}(\bm{\Sigma},m)$ has the Cholesky decomposition $\bm{T}\bm{T}^T=\bm{A}$. Also $\bm{A}=\bm{L}\bm{U}\bm{L}^T$ where $\bm{L}\bm{L}^T=\bm{\Sigma}$ and $\bm{U}\sim\mathcal{W}(\bm{I},m)$ with Cholesky decomposition $\bm{D}\bm{D}^T=\bm{U}$. If $\{t_{kl}\}$ and $\{d_{kl}\}$ are the Cholesky decomposition elements of $\bm{A}$ and $\bm{U}$, respectively, then
        \begin{enumerate}
            \item \textbf{Diagonal}:  $t_{kk}^2=d_{kk}^2l_{kk}^2\sim\mathcal{GA}(\frac{m-(k-1)}{2},2l_{kk}^2/m)$, where $l_{kk}$ is the $k$-th diagonal element of $\bm{L}$;\\
            \item \textbf{Off-Diagonal}: if we partition $\bm{L}$, $\bm{T}$, and $\bm{D}$ as
            \begin{equation}
            \bm{L}=\begin{bmatrix}
            \bm{L}_{11} & \bm{0}\\
            \bm{L}_{21} & \bm{L}_{22}
            \end{bmatrix}, \bm{T}=\begin{bmatrix}
            \bm{T}_{11} & \bm{0}\\
            \bm{T}_{21} & \bm{T}_{22}
            \end{bmatrix}, \bm{D}=\begin{bmatrix}
            \bm{D}_{11} & \bm{0}\\
            \bm{D}_{21} & \bm{D}_{22}
            \end{bmatrix}
            \end{equation}
            where $\bm{L}_{kk}$, $\bm{T}_{kk}$, $\bm{D}_{kk}$ are $p_k\times p_k$ square matrices, then $$\bm{T}_{21}\mid \bm{T}_{11}\sim\mathcal{MN}_{p_2,p_1}(\bm{L}_{21}\bm{D}_{11},\bm{L}_{22}\bm{L}_{22}^T/m,\bm{I}_{p_1}).$$
        \end{enumerate}
\end{corollary}

   \begin{proof}
$  $

\begin{enumerate}
\item \textbf{Diagonal}: It is easy to show $t_{kk}^2=d_{kk}^2l_{kk}^2$ and $d_{kk}^2\sim \mathcal{GA}(\frac{m-(k-1)}{2},2/m)$, hence $t_{kk}^2=d_{kk}^2l_{kk}^2\sim\mathcal{GA}(\frac{m-(k-1)}{2},2l_{kk}^2/m)$;\\
\item \textbf{Off-Diagonal}: From Theorem \ref{theorem:chol} and \citet[][Theorem 2.3.1]{gupta1999matrix}, we have $\bm{D}_{21}\mid \bm{D}_{11}\sim\mathcal{MN}_{p_2,p_1}(\bm{D}_{11},\bm{I}_{p_2},\bm{I}_{p_1})$. We know
\begin{equation}
\begin{bmatrix}
\bm{T}_{11} & \bm{0}\\
\bm{T}_{21} & \bm{T}_{22}
\end{bmatrix}=\begin{bmatrix}
\bm{L}_{11} & \bm{0}\\
\bm{L}_{21} & \bm{L}_{22}
\end{bmatrix}\begin{bmatrix}
\bm{D}_{11} & \bm{0}\\
\bm{D}_{21} & \bm{D}_{22}
\end{bmatrix}=\begin{bmatrix}
\bm{L}_{11}\bm{D}_{11} & \bm{0}\\
\bm{L}_{21}\bm{D}_{11}+\bm{L}_{22}\bm{D}_{21} & \bm{L}_{22}\bm{D}_{22}
\end{bmatrix}
\end{equation}
Given \citet[][Theorem 2.3.10]{gupta1999matrix}, we have $$\bm{T}_{21}\mid \bm{T}_{11}\sim\mathcal{MN}_{p_2,p_1}(\bm{L}_{21}\bm{D}_{11},\bm{L}_{22}\bm{L}_{22}^T/m,\bm{I}_{p_1}).$$
\end{enumerate}
Now, we have proven all the statements.
\end{proof}

By Corollary \ref{theorem:chol2}, the diagonal elements $t^2_{ikk}(\bm{s})$ marginally follow gamma distributions
\begin{equation}
e^2_{ikk}(\bm{s})\sim \mathcal{GA}\left(\frac{m-(k-1)}{2},\frac{2e^{2\bm{X}_i\bm{\beta}_{kk}(\bm{s})}}{m}\right).
\end{equation}

 The marginal distributions of off-diagonal elements are
\begin{equation}
\begin{aligned}
e_{{i21}}(\bm{s})\mid d_{{i11}}(\bm{s})&\sim \mathcal{N}(d_{{i11}}(\bm{s})\bm{X}_i\bm{\beta}_{21}(\bm{s}),e^{2\bm{X}_i\bm{\beta}_{22}(\bm{s})}/m)\\
e_{{i31}}(\bm{s})\mid d_{{i11}}(\bm{s}),d_{{i21}}(\bm{s})&\sim \mathcal{N}(d_{{i11}}(\bm{s})\bm{X}_i\bm{\beta}_{31}(\bm{s})+d_{i21}(\bm{s})\bm{X}_i\bm{\beta}_{32}(\bm{s}),e^{2\bm{X}_i\bm{\beta}_{33}(\bm{s})}/m)\\
t_{{i32}}(\bm{s})\mid d_{{i22}}(\bm{s})&\sim \mathcal{N}(d_{{i22}}(\bm{s})\bm{X}_i\bm{\beta}_{32}(\bm{s}),e^{2\bm{X}_i\bm{\beta}_{33}(\bm{s})}/m),
\end{aligned}
\end{equation}
where $\bm{Z}_{ij}(\bm{s})=[Z_{ij1}(\bm{s}),Z_{ij2}(\bm{s}),Z_{ij3}(\bm{s})]^T$, $\bm{Z}_{ij}\sim\mathcal{GP}(\bm{0}, \mathcal{K}(\bm{s},\bm{s}'\mid \bm{\Phi}_u), \bm{I})$ (the term is independent distributed over $j$ here, and $i$ is a given and fixed subject index), and $d_{ikl}(\bm{s})$ is the (k,l)-th Cholesky decomposition elements of $\sum_{j=1}^m\bm{Z}_{ij}(\bm{s})\bm{Z}_{ij}(\bm{s})^T/m$.

Now we give Theorem \ref{theorem:chol3}:
\begin{theorem}
    \label{theorem:chol3}
    If $\bm{U}\sim \mathcal{SWP}(m,\mathcal{K}(\bm{s},\bm{s}'\mid \bm{\Phi}),\bm{I})$, $\bm{Z}_j=[\bm{Z}_{j1}^T,\bm{Z}_{j2}^T]^T\sim\mathcal{GP}(\bm{0},\mathcal{K}(\bm{s},\bm{s}'\mid \bm{\Phi}),\bm{I})$, and $\{d_{{kl}}(\bm{s})\}$ are elements of $\bm{U}(\bm{s})$'s Cholesky decomposition with $\bm{D}(\bm{s})\bm{D}(\bm{s})^T=\bm{U}(\bm{s})$, then
        \begin{enumerate}
            \item \textbf{Diagonal}:  $d_{kk}^2\sim\mathcal{SWP}(m-(k-1),\mathcal{K}(\bm{s},\bm{s}'\mid \bm{\Phi}),1)$;\\
            \item \textbf{Off-diagonal}: If we partition $\bm{U}(\bm{s})$ and $\bm{D}(\bm{s})$ as
            $\bm{U}(\bm{s})=\begin{bmatrix}
            \bm{U}_{11}(\bm{s}) & \bm{U}_{12}(\bm{s})\\
            \bm{U}_{21}(\bm{s}) & \bm{U}_{22}(\bm{s})
            \end{bmatrix}$, $\bm{D}(\bm{s})=\begin{bmatrix}
            \bm{D}_{11}(\bm{s}) & \bm{0}\\
            \bm{D}_{21}(\bm{s}) & \bm{D}_{22}(\bm{s})
            \end{bmatrix}$, respectively, where $\bm{U}_{kk}(\bm{s})$ and $\bm{D}_{kk}(\bm{s})$ are $p_k\times p_k$ square matrices with $\bm{U}_{11}(\bm{s})=\frac{1}{m}\sum_{j =1}^{m}\bm{Z}_{j 1}(\bm{s})\bm{Z}_{j 1}(\bm{s})^T$,
            then conditional on $\mathbf{S}_{\bm{D}_{21}}=\{\bm{Z}_{j1}(\bm{s}) : j=\{1,2,...,m\}, \bm{s}\in\mathcal{D}\}$, the term  $\{vect[\bm{D}_{21}(\bm{s})]:\bm{s}\in\mathcal{D}\}$ is a mean-zero multivariate Gaussian process with spatial covariance ${cov}(vect[\bm{D}_{21}(\bm{s})],vect[\bm{D}_{21}(\bm{s}')]\mid \mathbf{S}_{\bm{D}_{21}})=\mathcal{K}(\bm{s},\bm{s}'\mid \bm{\Phi})\bm{I}_{p_2}\otimes\frac{1}{m}\bm{D}_{11}(\bm{s})^{-1}\left[\sum_{j =1}^{m}\bm{Z}_{j  1}(\bm{s})\bm{Z}_{j  1}(\bm{s}')^T/m\right][\bm{D}_{11}(\bm{s}')^{-1}]^{T}$. $vect$ is a notation for matrix vectorization. 
        \end{enumerate}
\end{theorem}

\begin{proof}
$  $

\begin{enumerate}
\item \textbf{Diagonal}:  \citet[][Theorem 3.4.4]{mardia1980multivariate}  have given the relevant proofs for the non-spatial case and it is not difficult to extend it to the spatial case. \citet[][Section 8.2]{eaton2008multivariate} give that $m\bm{U}_{22.1}(\bm{s})=m\bm{D}_{22}(\bm{s})\bm{D}_{22}(\bm{s})^T=\bm{X}_2(\bm{s})^T\bm{X}_2(\bm{s})-\bm{X}_2(\bm{s})^T\bm{X}_1(\bm{s})(\bm{X}_1(\bm{s})^T\bm{X}_1(\bm{s}))^{-1}\bm{X}_1(\bm{s})^T\bm{X}_2(\bm{s})=\bm{X}_2(\bm{s})^T\bm{Q}(\bm{s})\bm{X}_2(\bm{s})$. In \citet{mardia1980multivariate}, the term $\bm{X}_q(\bm{s})=[\bm{Z}_{1q}(\bm{s}),...,\bm{Z}_{mq}(\bm{s})]^T$ is called normal data matrix of $\mathcal{N}(\bm{0},\bm{I}_{p_q})$ (marginally). The term $\bm{Q}(\bm{s})$ is an idempotent matrix with rank $m-p_1$. We do spectral decomposition of $\bm{Q}(\bm{s})$ to have $\bm{Q}(\bm{s})=\bm{\Gamma}(\bm{s})^T\bm{\Lambda}\bm{\Gamma}(\bm{s})$ and the property of idempotent matrix gives that $\bm{\Lambda}$ is a diagonal matrix with $m-p_1$ non-zero eigenvalues which equal to 1. The proof of \citet[][Theorem 3.4.4]{mardia1980multivariate} shows that $\bm{Y}(\bm{s})=\bm{\Gamma}(\bm{s})\bm{X}_2(\bm{s})$ is a normal data matrix of $\mathcal{N}(\bm{0},\bm{I}_{p_2})$ given \citet[][Theorem 3.3.2]{mardia1980multivariate}. We have $cor(\bm{Y}(\bm{s}),\bm{Y}(\bm{s}')\mid \bm{\Gamma}(\bm{s}))=\mathcal{K}(\bm{s},\bm{s}')=cor(\bm{Y}(\bm{s}),\bm{Y}(\bm{s}'))$. Then we finally have $\bm{D}_{22}(\bm{s})\bm{D}_{22}(\bm{s})^T=\frac{1}{m}\sum_{j=1}^{m-p_1}\bm{Y}_j(\bm{s})\bm{Y}_j(\bm{s})^T$ and $\bm{Y}_j\sim\mathcal{GP}(0,\mathcal{K}(\bm{s},\bm{s}'\mid \bm{\Phi}),\bm{I})$. Therefore, we prove the first statement.\\
\item \textbf{Off-Diagonal}: 
We show the result using moment generating function. For a concise illustration, we give $p_1=p_2=1$.

\begin{equation}
    \begin{aligned}
    &M(t_{\bm{s}_1},...t_{\bm{s}_n})\\
    &=\mathbb{E}[\exp(t_{\bm{s}}d_{21}(\bm{s}_1)+...+t_{\bm{s}'}d_{21}(\bm{s}_n))\mid Z_{j1},j=\{1,2,...,m\}]\\
    &=\mathbb{E}[t_{\bm{s}}\frac{1}{m}\sum_{j =1}^{m}Z_{j  2}(\bm{s})Z_{j  1}(\bm{s})/d_{11}(\bm{s})+t_{\bm{s}'}\frac{1}{m}\sum_{j =1}^{m}Z_{j  2}(\bm{s}')Z_{j  1}(\bm{s}')/d_{11}(\bm{s}')\mid Z_{j1},j=\{1,2,...,m\}]\\
    &\text{because $\frac{1}{m}\sum_{j =1}^{m}Z_{j  2}(\bm{s})Z_{j  1}(\bm{s})=U_{21}(\bm{s})=d_{21}(\bm{s})d_{11}(\bm{s})$), see the proof of \citet[][Lemma 8.10]{eaton2008multivariate}}\\
    &=\exp([t_{\bm{s}_1},..., t_{\bm{s}_n}]^T\bm{Q}[t_{\bm{s}_1},..., t_{\bm{s}_n}])\\
    &\text{(because $Z_{j2}\sim \mathcal{GP}(0,\mathcal{K}(\bm{s},\bm{s}'\mid \bm{\Phi}),1)$)}
    \end{aligned}
\end{equation}
$\bm{Q}$ is a symmetric matrix with diagonal elements are $\frac{1}{m^2}\sum_{j =1}^{m}Z_{j  1}(\bm{s})^2$ for $\bm{s}\in\mathcal{D}$ and off-diagonal elements $\frac{1}{m^2}\sum_{j =1}^{m}Z_{j  1}(\bm{s})Z_{j  1}(\bm{s}')\mathcal{K}(\bm{s},\bm{s}'\mid \bm{\Phi})$ for all $\bm{s}\not=\bm{s}'$.

Then we conclude that the off-diagonal elements $d_{21}$ are conditionally normal distributed. It is not difficult to extend this result to multivariate/matrix-variate cases. Here, we just need to derive the covariance. Following the result that $\frac{1}{m}\sum_{j =1}^{m}\bm{Z}_{j  2}(\bm{s})\bm{Z}_{j  1}(\bm{s})^T=\bm{U}_{21}(\bm{s})=\bm{D}_{21}(\bm{s})\bm{D}_{11}(\bm{s})^T$, we have 
\begin{equation}
    \begin{aligned}
    &cov(vect[\bm{D}_{21}(\bm{s})],vect[\bm{D}_{21}(\bm{s}')]\mid \bm{S}_{\bm{D}_{21}})\\
    &=\mathcal{K}(\bm{s},\bm{s}'\mid \bm{\Phi})\bm{I}_{p_2}\otimes\frac{1}{m}\bm{D}_{11}(\bm{s})^{-1}\left[\sum_{j =1}^{m}\bm{Z}_{j  1}(\bm{s})\bm{Z}_{j  1}(\bm{s}')^T/m\right][\bm{D}_{11}(\bm{s}')^{-1}]^{T}
    \end{aligned}
\end{equation}
\end{enumerate}
Now, we have proven all the statements.
\end{proof}

Given Theorem \ref{theorem:chol3}, we have that the diagonal element $e^2_{ikk}$ follow a spatial Wishart process, denoted as
\begin{equation}
\label{eq:finaldiag}
    e_{ikk}^2\sim\mathcal{SWP}(m-(k-1), \mathcal{K}(\bm{s},\bm{s}'\mid \bm{\Phi}_u)e^{\sum_{\bm{\omega}\in\{\bm{s},\bm{s}'\}}2\bm{X}_i\bm{\beta}_{kk}(\bm{\omega})},1).
\end{equation}
In addition, the off-diagonal element $e_{ikl}$ conditionally follow a univariate spatial Gaussian process, denoted as
\begin{equation}
\label{eq:finaloff}
\begin{aligned}
&e_{{i21}}\mid \bm{S}_{e_{i21}}\sim\mathcal{GP}(d_{i11}(\bm{s})\bm{X}_i\bm{\beta}_{21}(\bm{s}),\mathcal{K}(\bm{s},\bm{s}'\mid \bm{\Phi}_u)Q_{ij1}\ e^{\sum_{\bm{w}\in \{\bm{s},\bm{s}'\}}\bm{X}_i\bm{\beta}_{22}(\bm{w})},\frac{1}{m})\\
&e_{{i31}}\mid \bm{S}_{e_{i31}}\sim\mathcal{GP}(d_{i11}(\bm{s})\bm{X}_i\bm{\beta}_{31}(\bm{s})+d_{{i21}}(\bm{s})\bm{X}_i\bm{\beta}_{32}(\bm{s}),\mathcal{K}(\bm{s},\bm{s}'\mid \bm{\Phi}_u)Q_{ij1}\ e^{\sum_{\bm{w}\in \{\bm{s},\bm{s}'\}}\bm{X}_i\bm{\beta}_{33}(\bm{w})},\frac{1}{m})\\
&e_{{i32}}\mid \bm{S}_{e_{i32}}\sim\mathcal{GP}(d_{i22}(\bm{s})\bm{X}_i\bm{\beta}_{32}(\bm{s}),\mathcal{K}(\bm{s},\bm{s}'\mid \bm{\Phi}_u)Q_{ij2}\ e^{\sum_{\bm{w}\in \{\bm{s},\bm{s}'\}}\bm{X}_i\bm{\beta}_{33}(\bm{w})},\frac{1}{m}),
\end{aligned}
\end{equation}
where $Q_{ijl}=\frac{\sum_{j =1}^{m}Z_{ij  l}(\bm{s})Z_{ij  l}(\bm{s}')/m}{d_{ill}(\bm{s})d_{ill}(\bm{s}')}$. Note that the conditional component $\bm{S}_{e_{ikl}}$ is invariant to the terms on the right-hand side.

\subsection*{Proofs of Asymptotic Results}
\begin{proof}[of Theorem~\ref{thm:eq}]

The spatial joint distribution of diagonal elements $e_{ikk}(\bm{s})$ is a multivariate Gamma distribution defined in \citet{krishnaiah1961remarks}. \citet[][Lemma 1]{krishnaiah1961remarks} also show that the correlation matrix of the Gamma random variables is a matrix defined by $\mathcal{C}(\bm{s},\bm{s}'\mid \bm{\Phi}_u)=\mathcal{K}(\bm{s},\bm{s}'\mid \bm{\Phi}_u)^2$. Since Gamma distribution can be considered as a summation of independent squared Gaussian distributed variables ($m$ is an integer), it is not difficult to have that as $m\rightarrow\infty$, we have the result of convergence in distribution:
$$m^{1/2}[e_{ikk}^2(\bm{s}_1)-e^{2\bm{X}_i\bm{\beta}_{kk}(\bm{s}_1)},...,e_{ikk}^2(\bm{s}_n)-e^{2\bm{X}_i\bm{\beta}_{kk}(\bm{s}_n})]^T\rightarrow\mathcal{GP}(0,\mathcal{C}(\bm{s},\bm{s}'\mid \bm{\Phi}_u)e^{\sum_{\bm{\omega}\in\{\bm{s},\bm{s}'\}}2\bm{X}_i\bm{\beta}_{kk}(\bm{\omega})},2),$$ given the central limit theorem and Slutsky's theorem. Taking the logarithm on the diagonal elements and applying the delta method, the term $e^{\sum_{\bm{\omega}\in\{\bm{s},\bm{s}'\}}2\bm{X}_i\bm{\beta}_{kk}(\bm{\omega})}$ is cancelled. We show the detailed steps below:

Let $g(X_1, ..., X_n)=[\frac{1}{2}\log X_1, ..., \frac{1}{2}\log X_n]^T$. We have $\nabla g(X_1, ..., X_n)=\frac{1}{2}diag[\frac{1}{X_1},...,\frac{1}{X_n}]$. We know the covariance matrix of $m^{1/2}[e_{ikk}^2(\bm{s}_1)-e^{2\bm{X}_i\bm{\beta}_{kk}(\bm{s}_1)},...,e_{ikk}^2(\bm{s}_n)-e^{2\bm{X}_i\bm{\beta}_{kk}(\bm{s}_n})]^T$ is $\bm{O}=2\times diag(e^{2\bm{X}_i\bm{\beta}_{kk}(\bm{s}_1)}, ..., e^{2\bm{X}_i\bm{\beta}_{kk}(\bm{s}_n)})\bm{R}_c diag(e^{2\bm{X}_i\bm{\beta}_{kk}(\bm{s}_1)}, ..., e^{2\bm{X}_i\bm{\beta}_{kk}(\bm{s}_n)})$, where $\bm{R}_c$ is the correlation matrix constructed by $\mathcal{C}(\bm{s},\bm{s}'|\bm{\Phi}_u)$. Following delta method, we have that the covariance matrix of $m^{1/2}[\log e_{ikk}(\bm{s}_1)-\bm{X}_i\bm{\beta}_{kk}(\bm{s}_1), ..., \log e_{ikk}(\bm{s}_n)-\bm{X}_i\bm{\beta}_{kk}(\bm{s}_n)]^T$ is $\mathbf{K}=\nabla g(X_1, ..., X_n)^T\mathbf{O}\nabla g(X_1, ..., X_n)=\frac{1}{2}\mathbf{R}_c$. Hence, we have the result of convergence in distribution: $$m^{1/2}[\log e_{ikk}(\bm{s}_1)-\bm{X}_i\bm{\beta}_{kk}(\bm{s}_1), ..., \log e_{ikk}(\bm{s}_n)-\bm{X}_i\bm{\beta}_{kk}(\bm{s}_n)]^T \rightarrow\mathcal{GP}\left(0,\mathcal{C}(\bm{s},\bm{s}'\mid \bm{\Phi}_u),{\frac{1}{2}}\right).$$
Therefore, we have proven the first statement. Please note here we abuse the notation  $\mathcal{GP}$ because this notation is originally created for a process. However, since it does not cause any problems when used for a distribution, we simply continue to use $\mathcal{GP}$ without introducing any additional notations.

We note that as $m\rightarrow\infty$, $d_{ikk}(\bm{s})$ converges in probability to $1$, $d_{ikl}(\bm{s})$ converges in probability to $0$, and $Q_{ijl}$ converges in probability to $\mathcal{K}(\bm{s},\bm{s}'\mid \bm{\Phi}_u)$, given law of large numbers and continuous mapping theorem \citep[Theorem 2.3]{van2000asymptotic}. We have $\mathbb{E}[[e_{{ikl}}(\bm{s}_1),...,e_{{ikl}}(\bm{s}_n)]^T\mid \bm{S}_{e_{ikl}}]$ converges in probability to $\mathbb{E}[[t_{{ikl}}(\bm{s}_1),...,t_{{ikl}}(\bm{s}_n)]^T\mid \bar{t}_{ikk}]$, $cor(e_{{ikl}}(\bm{s}),e_{{ikl}}(\bm{s}')\mid \bm{S}_{e_{ikl}})$ converges in probability to $\mathcal{K}(\bm{s},\bm{s}'\mid \bm{\Phi})^2=\mathcal{C}(\bm{s},\bm{s}'\mid \bm{\Phi})$.

Given the Slutsky's theorem and $\bm{\beta}_{kl}(\bm{s})=\bm{0}$, we have the following  results about convergence in distribution:

\begin{equation}
\label{eq:finaloff}
\begin{aligned}
&m^{1/2}e_{{i21}}\mid \bm{S}_{e_{i21}}\rightarrow\mathcal{GP}(0,\mathcal{C}(\bm{s},\bm{s}'\mid \bm{\Phi}_u)e^{\sum_{\bm{w}\in \{\bm{s},\bm{s}'\}}\bm{X}_i\bm{\beta}_{22}(\bm{w})},1)\\
&m^{1/2}e_{{i31}}\mid \bm{S}_{e_{i31}}\rightarrow\mathcal{GP}(0,\mathcal{C}(\bm{s},\bm{s}'\mid \bm{\Phi}_u)e^{\sum_{\bm{w}\in \{\bm{s},\bm{s}'\}}\bm{X}_i\bm{\beta}_{33}(\bm{w})},1)\\
&m^{1/2}e_{{i32}}\mid \bm{S}_{e_{i32}}\rightarrow\mathcal{GP}(0,\mathcal{C}(\bm{s},\bm{s}'\mid \bm{\Phi}_u) e^{\sum_{\bm{w}\in \{\bm{s},\bm{s}'\}}\bm{X}_i\bm{\beta}_{33}(\bm{w})},1),
\end{aligned}
\end{equation}
Also, we have $\bar{t}_{ikk}(\bm{s})$ converges in probability $e^{\bm{X}_i\bm{\beta}_{kk}(\bm{s})}$ as $N\rightarrow\infty$, given law of large number and continuous mapping theorem \citep[Theorem 2.3]{van2000asymptotic}. Given the Slutsky's theorem and  $\bm{\beta}_{kl}(\bm{s})=\bm{0}$, we have the following  results about convergence in distribution:

\begin{equation}
\label{eq:finaloff}
\begin{aligned}
&m^{1/2}t_{{i21}}\mid \bar{t}_{{i22}}\rightarrow\mathcal{GP}(0,\mathcal{C}(\bm{s},\bm{s}'\mid \bm{\Phi}_u)e^{\sum_{\bm{w}\in \{\bm{s},\bm{s}'\}}\bm{X}_i\bm{\beta}_{22}(\bm{w})},1)\\
&m^{1/2}t_{{i31}}\mid \bar{t}_{{i33}}\rightarrow\mathcal{GP}(0,\mathcal{C}(\bm{s},\bm{s}'\mid \bm{\Phi}_u)e^{\sum_{\bm{w}\in \{\bm{s},\bm{s}'\}}\bm{X}_i\bm{\beta}_{33}(\bm{w})},1)\\
&m^{1/2}t_{{i32}}\mid \bar{t}_{{i33}}\rightarrow\mathcal{GP}(0,\mathcal{C}(\bm{s},\bm{s}'\mid \bm{\Phi}_u) e^{\sum_{\bm{w}\in \{\bm{s},\bm{s}'\}}\bm{X}_i\bm{\beta}_{33}(\bm{w})},1).
\end{aligned}
\end{equation}

Now, we have proven the second statement because both converge to the same distribution/process.
\end{proof}

\subsection*{Cholesky Decomposition Process}
\begin{proof}[of Theorem~\ref{thm:valid}]

Since the measures on the Cholesky decomposition process model and the spatial Wishart process model can be one-to-one mutually transformed. It is not difficult to obtain the results based on the proofs in Appendix I. In addition, given the continuous mapping theorem \citep[Theorem 2.3]{van2000asymptotic}, we have the almost-sure continuity.
\end{proof}


\end{document}